\documentclass[twocolumn,times,trackchanges]{aastex63}
\usepackage[utf8]{inputenc}

\newcommand{\lahcomment}[1]{\textcolor{brown}{COMMENT: #1}}

\begin{document}
\title{Modeling the Multiwavelength Evolution of the V960 Mon System}
\author[0000-0002-9540-853X]{Adolfo Carvalho}
\affiliation{Department of Astronomy; California Institute of Technology; Pasadena, CA 91125, USA}
\author{Lynne A. Hillenbrand}
\affiliation{Department of Astronomy; California Institute of Technology; Pasadena, CA 91125, USA}
\author[0000-0003-0125-8700]{Franz-Josef Hambsch} 
\affiliation{Vereniging voor Sterrenkunde, werkgroep veranderlijke sterren, Oostmeers 122 C, B-8000 Brugge, Belgium}
\affiliation{Bundesdeutsche Arbeitsgemeinschaft f\"{u}r Ver\"{a}nderliche Sterne e.V. (BAV), Munsterdamm 90, D-12169 Berlin, Germany}
\affiliation{AAVSO, 49 Bay State Road, Cambridge, MA 02138, USA}
\author[0000-0001-9734-7374]{Shawn Dvorak}    
\affiliation{Rolling Hills Observatory, Lake County, Florida, USA}
\affiliation{AAVSO, 49 Bay State Road, Cambridge, MA 02138, USA}
\author[0000-0003-1799-1755]{Michael Sitko} 
\affiliation{Space Science Institute; Boulder, CO 80301, USA}
\author{Ray W. Russell}  
\affiliation{The Aerospace Corporation; Los Angeles, CA 90245, USA}   
\author{Victoria Hammond}  
\affiliation{Department of Physics; University of Cincinnati; Cincinnati, OH 45221, USA}
\affiliation{current address: University of Louisville Department of Surgery; Louisville, 550 S Jackson, KY 40202, USA}
\author[0000-0002-8293-1428]{Michael Connelley}
\affiliation{Institute for Astronomy, University of Hawaii at Manoa,
          640 N. Aohoku Place, Hilo, HI 96720, USA}
\author[0000-0003-1412-2028]{Michael C.B. Ashley}
\affiliation{School of Physics, University of New South Wales, Sydney, NSW 2052, Australia}
\author[0000-0001-9315-8437]{Matthew J. Hankins}
\affiliation{Arkansas Tech University, Russellville, AR 72801, USA}



\begin{abstract}

We study the evolution of the FU Ori object V960 Mon since its outburst, using available multi-wavelength photometric time series  over 8 years, complemented by several epochs of moderate-dispersion spectrophotometry. 
We find that the source fading can be well-described by a decrease in the temperature of the inner disk,
which results from a combination of decreasing accretion rate and increasing inner disk radius. 
We model the system with a disk atmosphere model that produces the observed variations 
in multi-band photometry (this paper) and high resolution spectral lines (a companion paper).

\end{abstract}


\section{Introduction}
FU Ori objects are a class of stellar variable consisting of disk-bearing young stellar objects (YSOs) that have undergone eruptive outbursts. The outbursts reach optical amplitudes of $\Delta V \sim 4-6$ mag \citep{herbig_eruptive_1977} and are attributed to a significant increase in the accretion rate of the disk onto the central star \citep{hartmann_fu_1996}. The relatively small sample of FU Ori objects seen to outburst show a variety of rise times and post-outburst fade timescales \citep{clarke2005}.

Studying the evolution of the spectral energy distribution (SED) of such objects during the development of the outburst and  also after its peak is critical to understanding the physics driving the outburst. High resolution spectra are also necessary to constrain system parameters independently of the broadband SED via the temperature and rotational broadening indicated by absorption lines. 

V960 Mon is one of the most well-studied objects to have undergone an FU Ori outburst. The outburst was reported in late 2014 \citep{Maehara_2014ATel.6770, Hillenbrand_2014ATel.6797, Reipurth_2015ATel.6862....1R,Hackstein_2014ATel.6838} and the object was followed in its post-outburst state by many photometric campaigns in the optical and near-infrared. The post-outburst fade of the object is remarkably short (as presented below in Figure~\ref{fig:AAVSO_lc}) 
relative to other objects like V1057 Cyg \citep{Szabo_V1057cyg_2021ApJ}.

\citet{kospal_progenitor_2015} gathered historical photometry of the target and found the progenitor SED is consistent with a 4000 K T-Tauri-like object. However, this assumed a distance to the target of 450 pc, whereas recent work has found a distance of 1120 pc \citep{kuhn_comparison_2019}. With the updated distance to the target, a single 4000 K young star is not luminous enough to account for the progenitor SED.  

A possible remedy to this is that the system may be at least a binary, or a triple. One faint companion at 0.2 arcsec separation and a candidate companion at 0.1 arcsec separation have been detected using adaptive optics imaging of the system \citep{CoG_companion_2015}. In addition, \citet{Hackstein_binary_2015} propose there is a close-in binary causing variation in the lightcurve at a 17 day period. However, this is inconsistent with the observed separations of either of the \citet{CoG_companion_2015} companions. 

Spectroscopic evolution of V960 Mon was discussed in \citet{park_high-resolution_2020} who presented
high dispersion optical data from the Bohyunsan Optical Echelle Spectrograph (BOES) 
and near infrared data from Immersion Grating Infrared Spectrograph (IGRINS). 

Here, we consider the photometric evolution of V960 Mon, complemented by several epochs of newly presented near-infrared moderate resolution spectra from IRTF/SpeX.
The post-outburst fade was also sampled by us spectroscopically with high-resolution optical spectra from Keck/HIRES and Lick/APF, which we present in a companion paper  (Carvalho et al. 2023b, in preparation), hereafter Paper II. 
That study will focus on the evolution of the high dispersion optical spectra and specifically the details of the line profile changes as the source has faded. In Paper II we also interrogate the evolving disk model presented below with respect to the spectroscopic evolution.

In this paper, we interpret the body of available photometric time series data 
in the context of an accretion disk model. In Section \ref{sec:all_data}, we discuss the photometric and spectroscopic data we use in our analysis. Section \ref{sec:modelFitting} presents our improvements to the pure accretion disk model described in \citet{Rodriguez_model_2022} and our modeling strategy for the V960 Mon system.  In Section \ref{sec:temp} we give evidence that disk cooling is driving the post-outburst fade, and provide evidence for the physical drivers of the cooling being a combination of decrease in accretion rate and increase in disk inner radius.  We address the possible presence of a passive disk component in Section \ref{sec:NIR}.  We then discuss our results in the context of recent literature, and provide further interpretation of our analysis in Section \ref{sec:discussion}. Our conclusions are briefly summarized in Section \ref{sec:conclusion}. 

\begin{figure*}[!htb]
    \centering
    \includegraphics[width=0.98\linewidth]{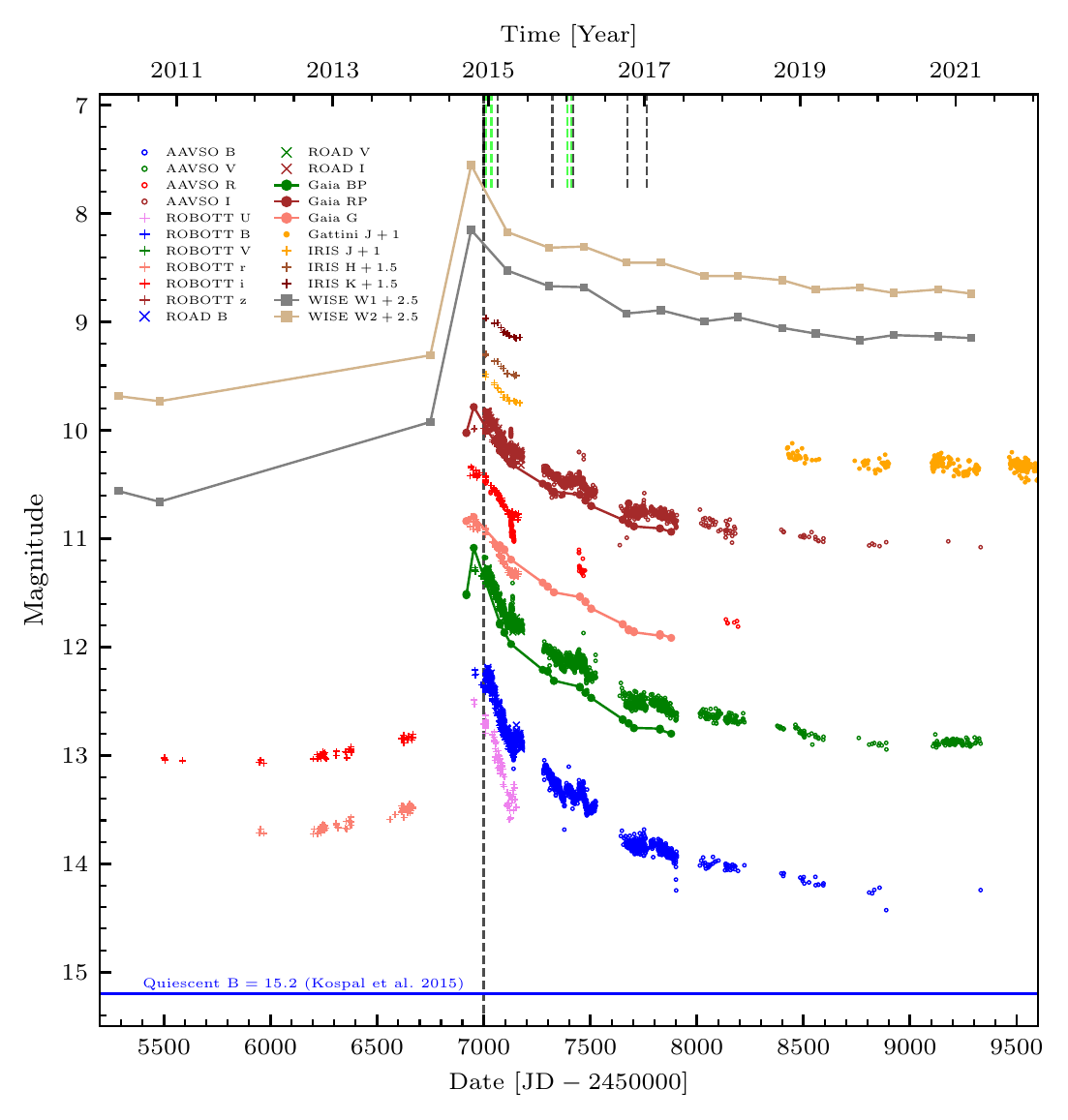}
    \caption{The multiband lightcurve of V960 Mon, showing photometry from many sources spanning $U$ band to the W1 and W2 filters of WISE/NEOWISE. The outburst and rapid fade of the target is clear in all bands, with the amplitude tending to decrease toward the red. Photometry that is more sparsely sampled (WISE, Gaia) is shown connected by lines to highlight the trend in the outburst and exponential fade. The black and lime vertical dashes mark the dates on which the HIRES and SpeX spectra, respectively, were collected. }
    \label{fig:AAVSO_lc}
\end{figure*}

\section{Data} \label{sec:all_data}
\subsection{Photometry} \label{sec:phot}
We use numerous archival sources and lightcurves assembled in previous literature of the target to study its post-outburst evolution across multiple bands spanning $U$ in the near-ultraviolet to WISE/W2 in the mid-infrared. 

\subsubsection{Visible Range}
The first 6 months post-outburst is well-sampled by the photometry assembled by and reported in \citet{Hackstein_binary_2015}. We use their data from Universitätssternwarte Bochum (USB) and The Remote Observatory Atacama Desert (ROAD). The data from USB were taken with RoBoTT ($U, B, V, r, i ,z$) are labelled as ROBOTT in our plots.   

We additionally use B, V, R, and I band lightcurves accessible in the American Association of Variable Star Observers (AAVSO) archive. The tight sampling of the data made it possible to follow the evolution of the closely during the early stages of the fade. To mitigate some of scatter in the photometry and enable better interpolation to observation dates of the spectroscopic data, we bin the AAVSO photometry in 10 day increments. 

We also include photometry from the Gaia mission, which we accessed via the Gaia ESA Archive \citep{Gaia_edr3_2021A&A, gaia_photometry_edr3_2021A&A}. The Gaia $BP$, $RP$, and $G$ band photometry show the outburst peak itself, in the MJD 56954 epoch. The Gaia photometry constrains the date and amplitude of the outburst peak and informs our interpretation of our other data on the system. It is especially important in our analysis below to confirm that the earliest epoch in the high dispersion spectra is representative of the outburst peak of the system.

\subsubsection{Infrared}
We assemble the infrared lightcurves from several sources. Monitoring of the target in $J,H,K$ bands 
was also done by \citet{Hackstein_binary_2015} 
using the 0.8 m Infrared Imaging System \citep[IRIS, ][]{Hodapp_IRIS_2010SPIE}. 

The Gattini infrared survey \citep{Moore_gattini_2016SPIE} provides J-band photometry of V960 Mon starting in September 2018, sampling the brightness of the target late along its plateau.

The object was also observed pre-outburst by the Wide-field Infrared Suvey Explorer (WISE) survey and post-outburst by the reactivation of the WISE mission as NEOWISE \citep{Mainzer_neowise_2011ApJ} which observes any given position approximately every six months. We use the $W1$ and $W2$ photometry, binning the point spread function fitting photometry from each visit to one point at the median epoch, median flux, and with error equivalent to the dispersion in the individual measurements. We perform the saturation correction provided in \citet{cutri_neowise_supplement_2015nwis} for the saturated observations following outburst. 

There is a nearby source ($\sim 5^{\prime \prime}$ to the North), but the FU Ori object is well-resolved from it in all photometry included in this work. The full lightcurve, including all data described above, is shown in Figure \ref{fig:AAVSO_lc}. We include the pre-outburst $B$ magnitude of the target reported by \citet{kospal_progenitor_2015} from 1953 and 1983 Palomar plates as a reference. 

The photometry corresponding to the epoch of our high-dispersion spectrum shortly following the outburst (marked in Figure \ref{fig:AAVSO_lc}) is used to produce an SED of the target. The SED is used to fit our disk model (see Section \ref{sec:modelFitting} for the SED and the disk model-fitting procedure).

\subsubsection{Multi-Wavelength Lightcurve and Photometric Evolution}

Figure \ref{fig:AAVSO_lc} shows the multi-band time-series photometry (lightcurve) of the target. Although V960 Mon has faded considerably from its peak brightness, the lightcurve shows a plateau from MJD 58500 to the present that is well above quiescence, indicating that the object is still accreting at a significant rate and has not yet fully returned to its pre-outburst low-state accretion rate.

The initial outburst spanned multiple magnitudes in every band shown in Figure \ref{fig:AAVSO_lc}. Using the quiescent state estimated by \citet{kospal_progenitor_2015}, the $\Delta B$ of the outburst was approximately 3 magnitudes\footnote{See also the historic lightcurve from \citet{Sepic_V960MonHarvardPlates_2016NewA}, constructed using Harvard photograpic plates, showing no previous evidence of eruptions from 1899 to 1989.}. The ROBOTT $r$ and $i$ photometry shows outburst amplitudes of 2.8 and 2.6 magnitudes, respectively. Near-infrared
$J$ band photometry from the 2-micron All Sky Survey \citep[2MASS,][]{cutri_neowise_supplement_2015nwis} compared with IRIS $J$ band photometry shows the object jumped by $\Delta J \sim 2.5$ mag. 
Even in the mid-infrared, the outburst amplitudes reach 2.5 and 2.2 for W1 and W2, respectively.  
The post-outburst fade measured by AAVSO $B$-band photometry shows that V960 Mon is still $\sim 1$ magnitude brighter than at quiescence, while at J-band the Gattini photometry indicate the object is 1.6 mag brighter than at quiescence, 
and similarly in NEOWISE photometry the plateau is about 1.5 mag brighter than the faint-state WISE measurement.

\subsection{Spectroscopy} \label{sec:data_spec}


Our disk models described in Section \ref{sec:modelFitting} are constrained by the photometry described above and by high and medium dispersion spectra.
\subsubsection{Optical}
We use a high dispersion visible range spectrum taken at the time of the object's outburst at the Keck Observatory using the HIgh Resolution Echelle Spectrograph \citep[HIRES;][]{Vogt1994}. The spectrum covers 4780-9220 \AA\ at a signal-to-noise of 170 at 7100 \AA\and was processed with the 2008 version of the MAKEE pipeline reduction package written by Tom Barlow\footnote{ {\url{https://sites.astro.caltech.edu/~tb/makee/}}}.   %

\begin{figure*}[!htb]
    \centering
    \includegraphics[width=0.98\linewidth]{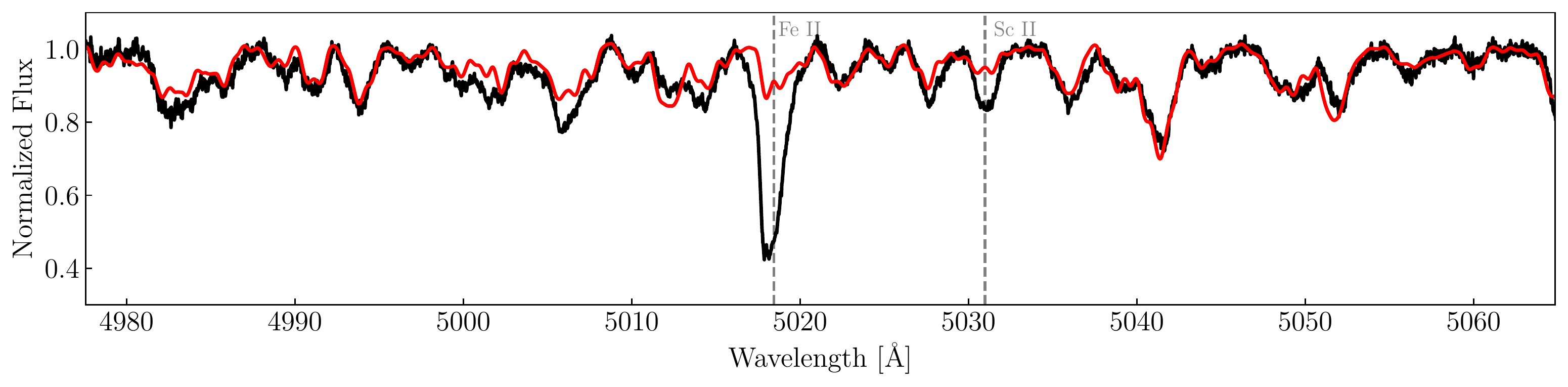}
    \includegraphics[width=0.98\linewidth]{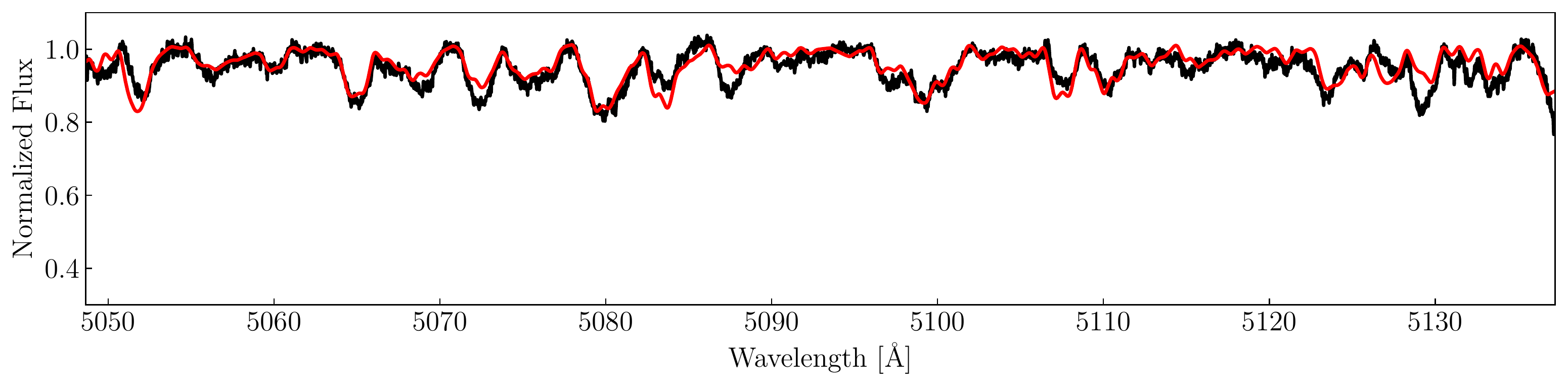}
    \includegraphics[width=0.98\linewidth]{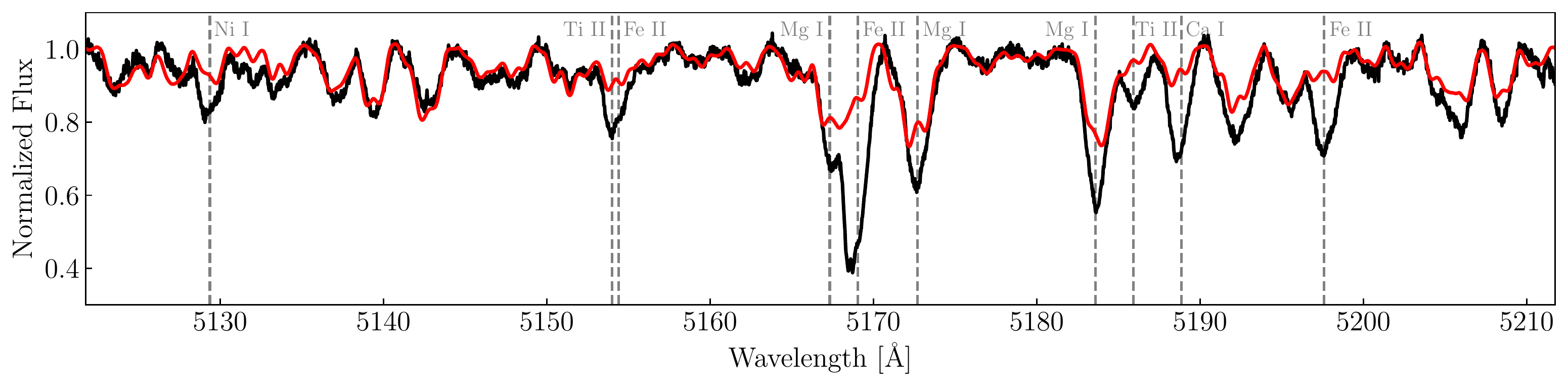}
    \caption{Orders of the HIRES spectrum taken closest to outburst that are used to compute the $\chi^2$ surface described in Section \ref{sec:modelFitting}. The fiducial model for the outburst spectrum is shown in red. Large mismatches between the data and model are seen in certain spectral lines; the extra depth (and blueshift in particular \ion{Fe}{2} lines) in the data is due to wind absorption, which is not included in our disk model, but will be the subject of a future paper.}
    \label{fig:hires_fits1}
\end{figure*}

\begin{figure*}[!htb]
    \centering
    \includegraphics[width=0.98\linewidth]{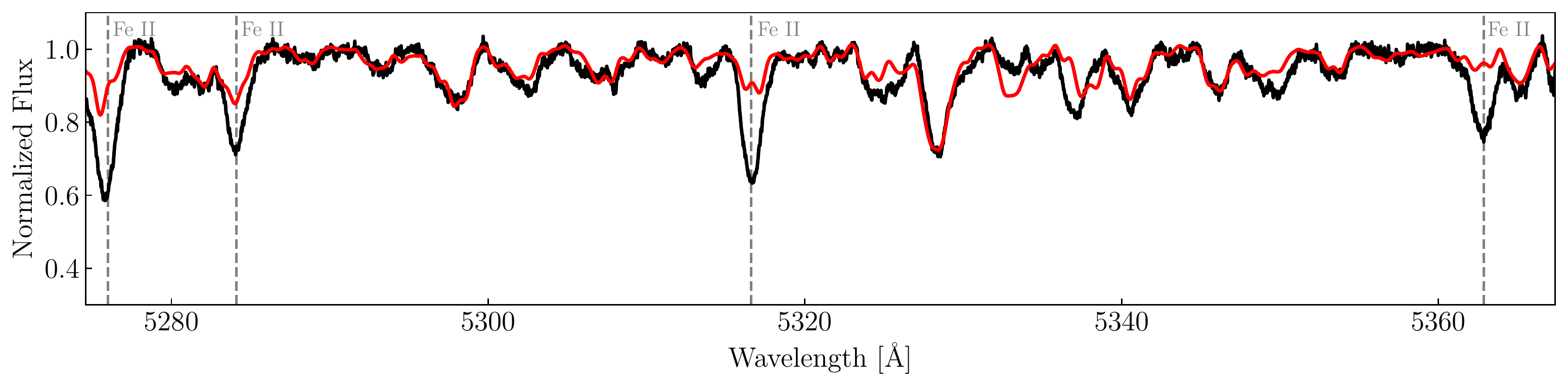}
    \includegraphics[width=0.98\linewidth]{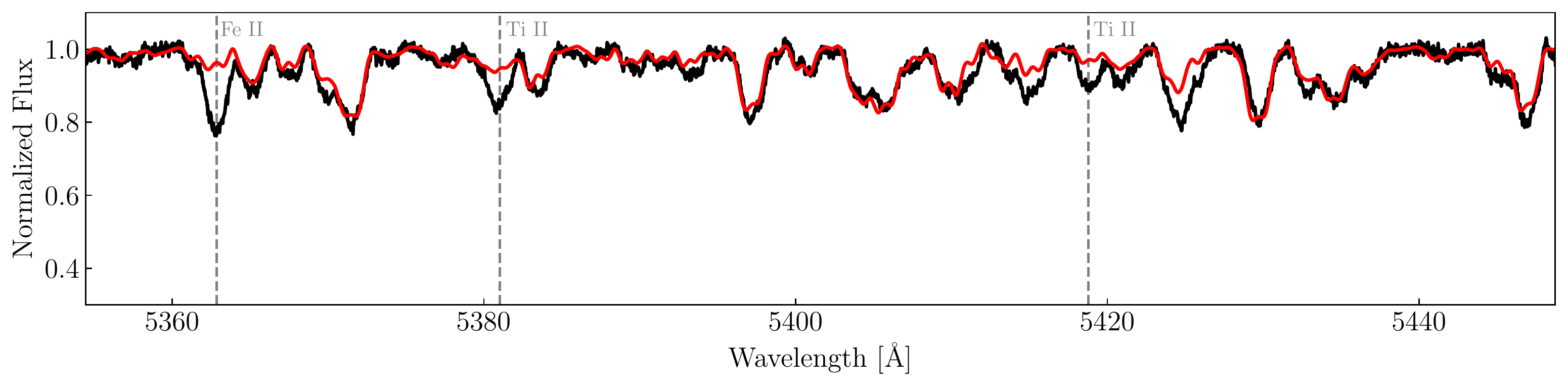}
    \includegraphics[width=0.98\linewidth]{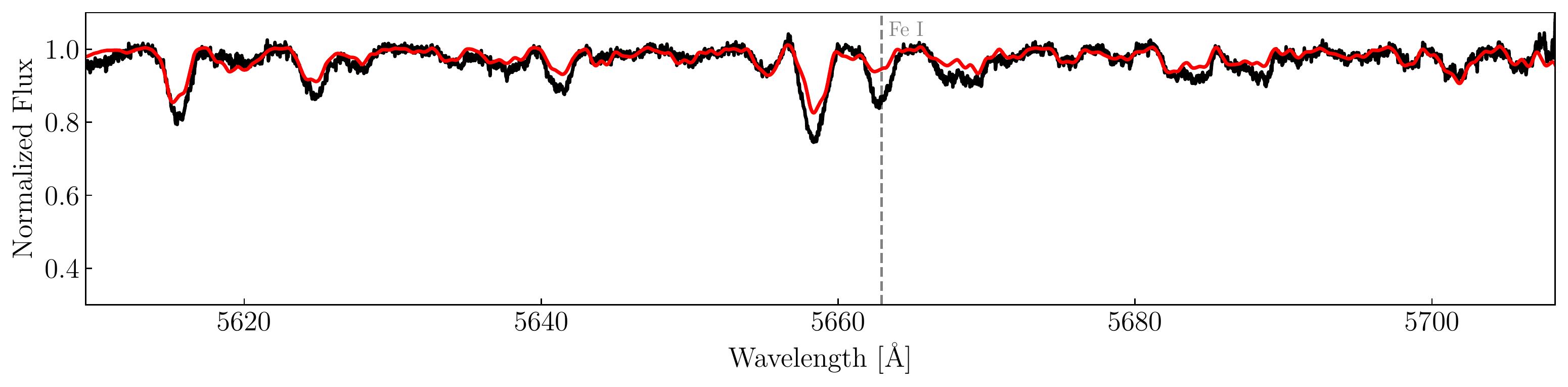}
    \caption{Same as Figure \ref{fig:hires_fits1} for the next three orders used.} 
    \label{fig:hires_fits2}
\end{figure*}

\begin{figure*}[!htb]
    \centering
    \includegraphics[width=0.98\linewidth]{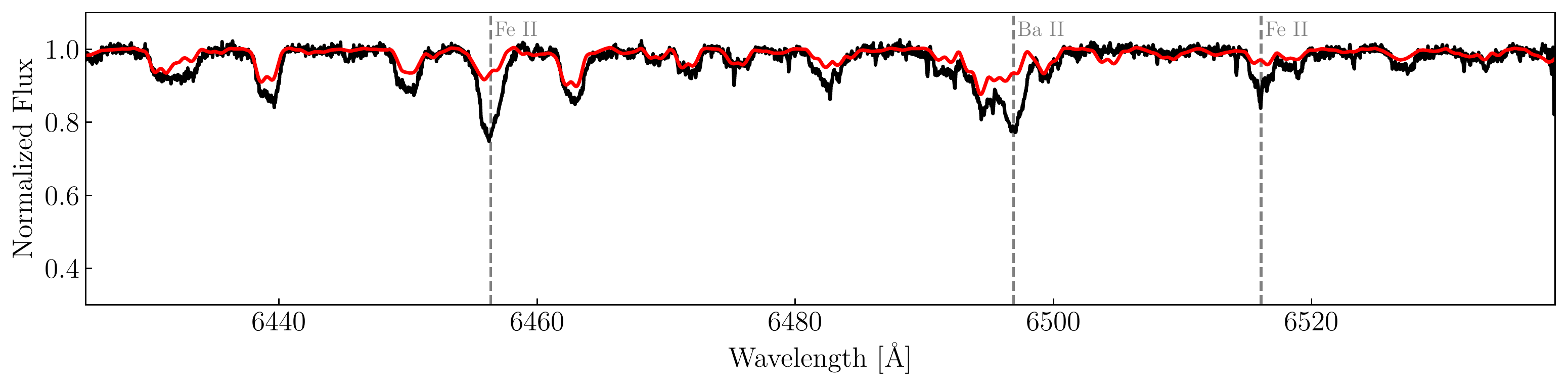}
    \includegraphics[width=0.98\linewidth]{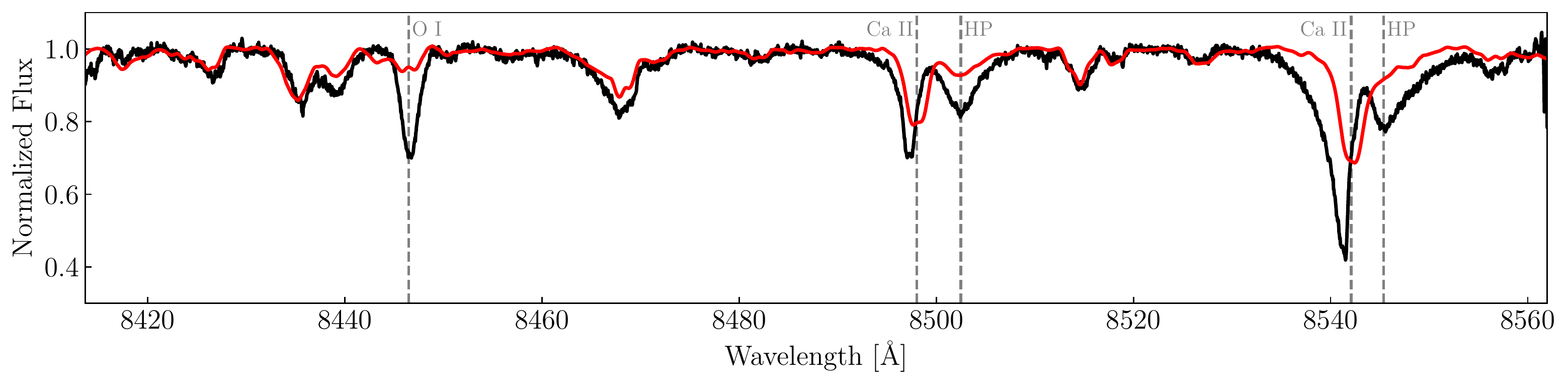}
    \includegraphics[width=0.98\linewidth]{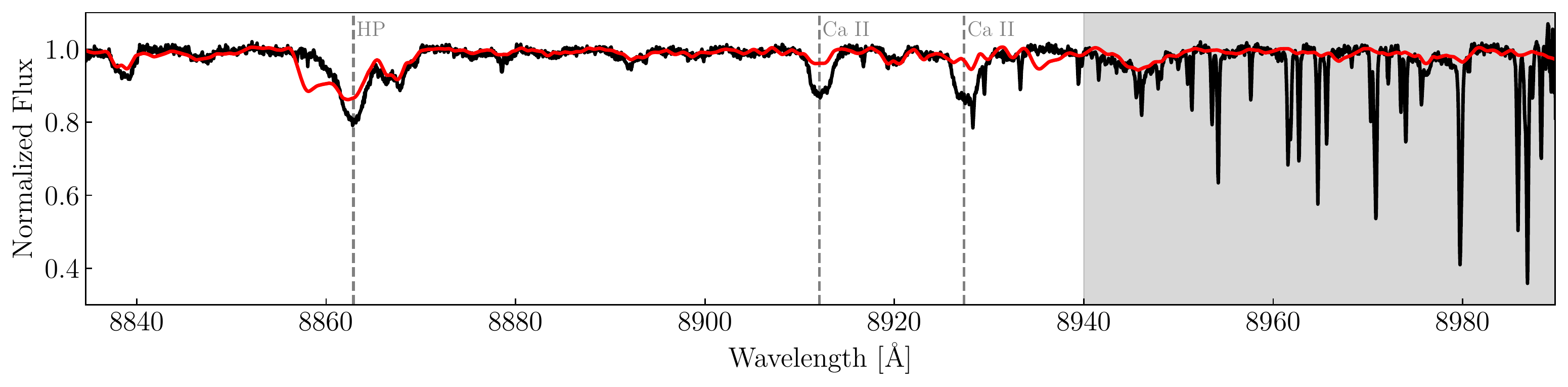}
    \caption{Same as Figure \ref{fig:hires_fits1} for three reddest orders used. Here, in addition to wind absorption that is not included in the model, there is also uncorrected telluric absorption in the data that appears as narrow lines toward the right of the bottom panel (grey shaded region).}
    \label{fig:hires_fits3}
\end{figure*}

The orders used in our accretion disk model fits are shown in Figures \ref{fig:hires_fits1}, \ref{fig:hires_fits2}, and \ref{fig:hires_fits3}. 
This spectrum is a part of a dedicated monitoring campaign which samples the post-outburst fade over several epochs, as indicated in Figure \ref{fig:AAVSO_lc} and given in Table~\ref{tab:SysParams}.  The spectra from this campaign are discussed in detail in Paper II. 




In the outburst spectrum, we compute the half-width-half-depth (HWHD) of several absorption lines across the optical range, and find that there is no correlation between wavelength of the line and HWHD.  The measurements are discussed and shown in detail in Paper II.  The mean and standard deviation of the measurements are $44 \pm 5$ km s$^{-1}$, consistent with the line-width measurements reported by \citet{park_high-resolution_2020}. As described below, this linewidth provides important constraints in the disk model that we explore for matching the lightcurve evolution.


We also measure a heliocentric system velocity of -43 km s$^{-1}$. This was determined by comparison of the spectrum with our high dispersion disk model, using it as a radial velocity reference. Our RV value is consistent with the system velocity of -42.6 km s$^{-1}$ found for CO by \citet[][submitted]{CruzSaenz_Apex_2023arXiv230103387C}, which is the RV value we adopt. 

\subsubsection{Infrared}
Our analysis also makes use of data from 
the NASA InfraRed Telescope Facility (IRTF) SpeX spectrograph \citep{Rayner_spex_2003PASP}.
We include the outburst epoch SpeX spectrum of V960 Mon obtained on 19 Dec 2014 and published by \citet{connelley_near-infrared_2018} who provide a detailed description of the data reduction.
We also make use of SpeX data taken at several other epochs: 15 Jan 2015, 6 Jan 2016, and 24 Jan 2016, all of which are previously unpublished. 

The SpeX observations were obtained using the cross-dispersed echelle gratings between 0.8 and 5.4 $\mu$m, using two spectrograph slit widths: 0.8$^{\prime \prime}$ (15 Jan 2015 and 24 Jan 2016) and 0.5$^{\prime \prime}$ (19 Dec 2014 and 6 Jan 2016). 
Spectral extraction, wavelength calibration, telluric correction, and flux calibration were done using the Spextool reduction package \citep{Vacca_telluricMethod_2003PASP, Cushing_spextool_2004PASP} running under IDL. $K$ band photometry was also obtained on each night using the SpeX guide camera, and extracted using aperture photometry. Due to the use of a slit with variations in throughput from seeing and tracking changes, all SpeX spectra were normalized to the $K$ band photometry.

The spectrophotometry from SpeX is an important supplement to the photometry when evaluating the success of our cooling accretion disk model.

\section{Disk Model and Fitting} \label{sec:modelFitting}
To constrain the physical parameters of the evolving V960 Mon system, we model the disk photosphere using a modified version of the model described in \citet{Rodriguez_model_2022}.  Our modifications to this model are described below. 
The model is based on the \citet{Shakura_sunyaev_alpha_1973A&A} viscous accretion disk radial temperature profile, 
\begin{equation} \label{eq:TProf}
    T^4_\mathrm{eff}(r) = \frac{3 G M_* \dot{M}}{8 \pi \sigma r^3} \left( 1 - \sqrt{\frac{R_\mathrm{inner}}{r}}  \right)    ,
\end{equation}
where $R_\mathrm{inner}$ is the inner radius of the accretion disk, $M_*$ is the mass of the central star, $\dot{M}$ is the stellar accretion rate, $G$ is the gravitational constant, and $\sigma$ is the Stefan-Boltzmann constant.

In this profile, a maximum temperature ($T_\mathrm{max}$) is reached at $r = 1.361 \ R_\mathrm{inner}$. Interior to this radius, the temperature falls rapidly to 0. This is not physical for this system, where we assume $R_\mathrm{inner} \sim R_*$ \citep[see, e.g., ][]{hartmann_fu_1996}. We therefore follow the example of \citet{Kenyon_FUOri_disks_1988ApJ} and set $T_\mathrm{eff}(r < 1.361 \ R_\mathrm{inner}) = T_\mathrm{max}$. 
The annuli at each $r$ value are then constructed from the appropriate PHOENIX \citep{Husser_Phoenix_2013A&A} model using the calculated $T_\mathrm{eff}(r)$. 

We fit the outburst SED of the target and the outburst HIRES spectrum using the modified model. We then use the photometry of the target to guide our modeling of the time-evolution of the system, which includes the evolution. The outburst epoch fitting procedure is described below, and the construction of models for later epochs is described in Section \ref{sec:colorTemp}.

\subsection{Improvements to the Rodriguez et al. 2022 Model}

We have implemented several improvements to increase the accuracy of the model and its ability to rapidly compute high dispersion spectra. One important improvement is to the rotational broadening calculation. \citet{Rodriguez_model_2022} use the popular method of rotationally broadening spectra by convolution with the kernel
\begin{equation}
    \phi (\lambda) = \left[ 1 - \left( \frac{\lambda - \lambda_0}{\lambda_{\mathrm{max}}(r)} \right)^2  \right]^{-\frac{1}{2}},
\end{equation}
where $\lambda_{\mathrm{max}}(r) = \lambda_0 \ \frac{v_{\mathrm{kep}}(r)}{c} \ \mathrm{sin} \ i$, $v_\mathrm{kep}(r)$ is the Keplerian velocity of the annulus centered at $r$, $i$ is the disk inclination, and $c$ is the speed of light in a vacuum. 
This method is only valid in small wavelength ranges ($\sim 100$ \AA). Rotationally broadening a broadband high dispersion spectrum with this method requires convolving dozens of 100 \AA\ segments at a time. This is extremely slow, and a different approach is needed to compute high dispersion models for the HIRES wavelength range of 4500 \AA\ to 9000 \AA\ quickly enough to reasonably use them for fitting. 

 We compute our rotational broadening by numerically integrating the disk 
 \begin{equation}
     F_\mathrm{broad}(r, \lambda) = \int^{2\pi}_0 F(r, \lambda(\theta)) d\theta
 \end{equation}
where
\begin{equation}
    \lambda(\theta) = \lambda \times \left( 1 + \frac{v_\mathrm{kep}(r)}{c}\mathrm{sin}( \theta) \  \mathrm{sin}(i) \right)    ,
\end{equation}
and $\theta$ is the azimuthal angle in the disk. 

This greatly accelerates the model calculation because we can integrate the spectrum in one operation, rather than folding it into 100 \AA\ chunks. 
(see \cite{Carvalho_RNAAS_rotbroad_2023RNAAS} for further detail).
The accelerated method enables us to rapidly produce high resolution model spectra, limited only by the large amount of memory required for the atmosphere model grid. 

Another significant improvement to the model is the calculation of log $g$ at each annulus in the disk. For this calculation, we assume the gravity is dominated by the central star and neglect self-gravity from the disk, so that the acceleration $g$ at an annulus with average radius $r$ is given by
\begin{equation}
    g(r) = \frac{G \ M_*}{r^2}
\end{equation}
where $G$ is the gravitational constant and $M_*$ is the mass of the central star.

The prescription for the disk gravity given in \citet{Chiang_Goldreich_passive_1997} is $g \sim \Omega^2 z$, where $\Omega$ is the Keplerian angular velocity and $z$ is the height of the disk atmosphere above the plane. Using $\Omega(r) = \sqrt{G M_*/r^3}$, we get $g \sim \frac{G M_*}{r^3} z$. If we assume that the disk flaring is constant and write $\alpha = \frac{z}{r}$, then $g \sim \alpha \frac{G M_*}{r^2}$, or $g \sim \alpha g_*$, where $g_*$ is the gravitational acceleration due only to the star. For thin disks, our approximation greatly overestimates the gravity. However, as flaring increases, our approximation becomes more accurate. The inner disk model presented in \citet{Zhu_outburst_FUOri_2020MNRAS} shows the flaring in the inner disk during an outburst may be as high as $z/r \sim 0.5-1$.

\subsection{Modeling Strategy and Results}
Our modeling strategy requires a broadband SED of the FU Ori object and a contemporaneous high resolution spectrum. In the case of V960 Mon, we construct the SED from the photometry shown in Figure \ref{fig:AAVSO_lc} near the peak, and use the outburst HIRES spectrum from the 2014 December 9 epoch. 

Using the two sets of data, we can constrain the mass of the central object, $M_*$, the innermost radius of the accretion disk, $R_\mathrm{inner}$, the inclination of the target, $i$, the accretion rate, $\dot{M}$, and the extinction to the target, $A_V$. 

Directly fitting the high dispersion spectrum alone proved challenging for a variety of reasons. One in particular is that there are broad combinations of the model parameters which produce fits with similar errors. The error surface then has many deep, steep-walled local minima that are difficult for optimization algorithms to escape from to find the global minimum.

Fitting the SED alone is not as significant a challenge, but the SED is quite limited in its ability to constrain many physical parameters simultaneously. The inclination of the system is degenerate with the overall luminosity, and the extinction to the system, $A_V$, is degenerate with the maximum temperature in the disk. These degeneracies result in broad posteriors in our SED fits when using uniform priors on the physical parameters. 

Combining the data to fit both the SED and high dispersion spectrum simultaneously is also challenging because the data volumes and ranges are vastly different. Defining properly scaled error terms in the combined space is unclear and requires further study.

We therefore adopt a hybrid approach: using a course sampling of the $\chi^2$ surface of the high dispersion data to inform priors on the parameters of the SED fit. This way, we impose both physically motivated and data-driven constraints on the most difficult-to-disentagle parameters such as $M_*$ and $\dot{M}$, which are highly covariant in the SED. 

\subsubsection{Fitting the High Dispersion Spectrum}

Before fitting the high dispersion spectrum, we compute several high dispersion models, spanning a broad range of $i$, $M_*$, $\dot{M}$, and $R_\mathrm{inner}$ parameter space. The widths of the absorption lines (see the empirical HWHD measurement above, and the $v \ \sin i \sim 60$ km s$^{-1}$ inference discussed in Paper II) are much smaller than the Keplerian velocities expected near the surface of young stars \citep[$150-300$ km s$^{-1}$, assuming $M_*$, $R_*$ values from 0.5 Myr PARSEC isochrones, ][]{Nguyen_parsec_2022A&A}. This immediately implies that the system is relatively face-on, with $i < 30^\circ$.

Varying $T_\mathrm{max}$, by choosing different values of $M_* \dot{M}$ and $R_\mathrm{inner}$, and comparing with the observed outburst spectrum allows us to estimate a reasonable $T_\mathrm{max}$ range of $6500 \ \mathrm{K} < T_\mathrm{max} < 8000 \ \mathrm{K}$. Assuming then reasonable $R_\mathrm{inner}$ values based on the $R_*$ values in the 0.5 Myr PARSEC isochrones \citep{Nguyen_parsec_2022A&A} for $0.1 \ M_\odot < M_* < 2 \ M_\odot$, gives a range of $1.5 \ R_\odot < R_\mathrm{inner} < 5 \ R_\odot$.  

We fix the system inclination to $15^\circ$, the best-fit value from the radiative transfer modeling in \citet{Kospal_ALMA_2021}. This choice of inclination agrees well with our expectation that the system is relatively face-on, despite the broad error bars reported in \citet{Kospal_ALMA_2021}. We continuum-normalize the models following the same normalization procedure applied to the HIRES spectra (see Paper II), eliminating $A_V$ as a parameter for the $\chi^2$ analysis.

Having established a reasonable range of $M_*$, $\dot{M}$, and $R_\mathrm{inner}$ values to explore, and fixing the other model parameters, we compute a grid of high dispersion models. We explore a range of $0.3 \ M_\odot < M_* < 1.4 \ M_\odot$, $1.5 \ R_\odot < R_\mathrm{inner} < 2.9 \ R_\odot$, and $-5 < \log \frac{\dot{M}}{M_\odot \ \mathrm{yr}^{-1}} < -4$ in steps of 0.1 $M_\odot$, 0.1 $R_\odot$, and 0.1 dex, respectively.
For each of the high-dispersion models, we compute the $\chi^2$ value. 

\begin{figure}[!htb]
    \centering
    \includegraphics[width=0.98\linewidth]{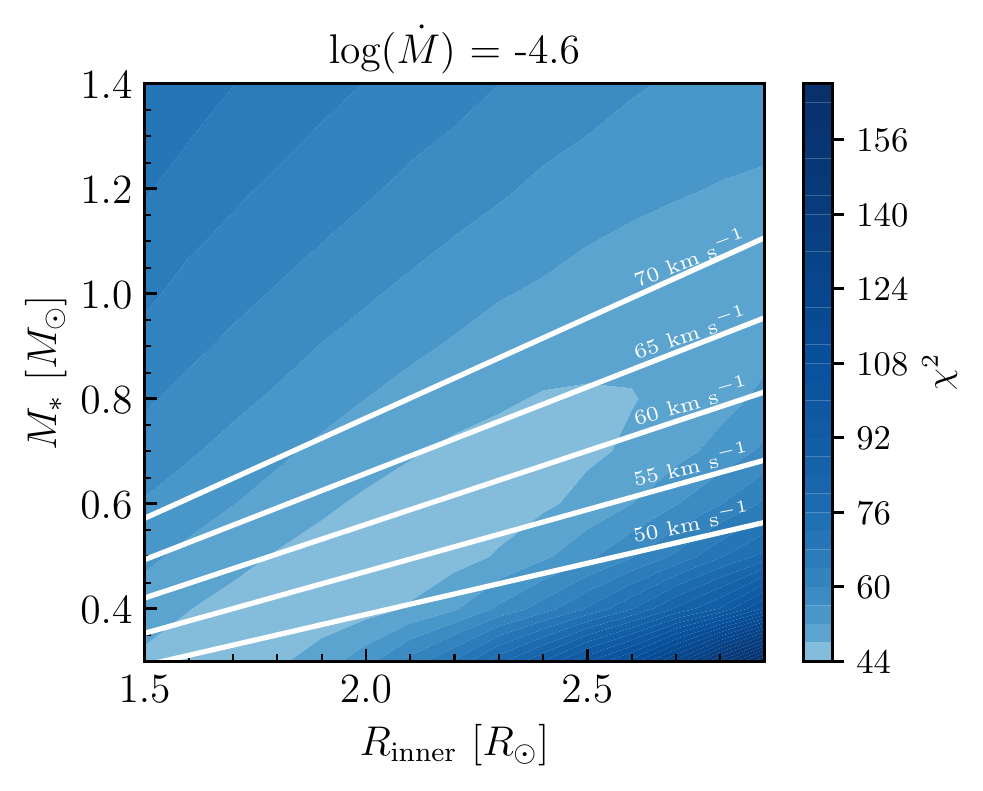}
    \caption{The $\chi^2$ surface for the outburst spectrum and high resolution models with varying $M_*$ and $R_{\text{inner}}$ values, log$(\dot{M})=-4.6$ $M_\odot$ yr$^{-1}$ and $i=15^{\circ}$. The mininmum $\chi^2$ value is found in the region that is intersected by the $v_{\text{max}}$sin $i$ = 60 km s$^{-1}$. The width of the minimum allows for an uncertainty of 5 km s$^{-1}$ on that value, which we use to constrain the SED fits.}
    \label{fig:ChisqGrid}
\end{figure}

We inspect the slices of the $\chi^2$ cube along the $\dot{M}$ dimension and find the $\dot{M} = 10^{-4.6}$ $M_\odot \ \mathrm{yr}^{-1}$ contains a minimum with physically reasonable values of $M_*$ and $R_\mathrm{inner}$ that are also consistent with the estimated $v_\mathrm{max} \sin i \sim 60$ km s$^{-1}$. The slice is shown in Figure \ref{fig:ChisqGrid}. For reference, we draw lines of constant $v_\mathrm{max} \ \mathrm{sin} \ i = v_\mathrm{kep} (R_\mathrm{inner}) \ \mathrm{sin} \ i$. 

Notice the valley of minimum $\chi^2$ values approximately intersects the 60 km s$^{-1}$ line at $M_* = 0.6 \ M_\odot$ and $R_\mathrm{inner} = 2.2 \ \mathrm{R_\odot}$. We adopt these as our best-fit parameters from the $\chi^2$ fit and will use them to construct the priors for the MCMC fits to the SED (Section \ref{sec:SEDMCMC}).

\subsubsection{Fitting the SED} \label{sec:SEDMCMC}

The first prior we impose on the SED fits is that the $M_*$ should be drawn from the \citet{Kroupa_IMF_review2001} Initial Mass Function (IMF), to properly reflect the lower probability that $M_*$ would be large.

We then use the estimated $v_\mathrm{max} \sin i \sim 60$ km s$^{-1}$ from the high dispersion data to construct a joint prior on $M_*$ and $R_\mathrm{inner}$. At each iterative step, we compute $v_\mathrm{max} \sin i = \sqrt{G M_*/R_\mathrm{inner}} \sin i$. We then impose that the resulting value be drawn from a Normal distribution, with a mean of $\bar{v} = 60$ km s$^{-1}$ and standard deviation of $\sigma_v = 5$ km s$^{-1}$. The $\sigma_v$ value is chosen from the approximate width of the valley in the $\chi^2$ plane, and thus represents the uncertainty on the best-fit parameters.  

\begin{figure}[!htb]
    \centering
    \includegraphics[width=\linewidth]{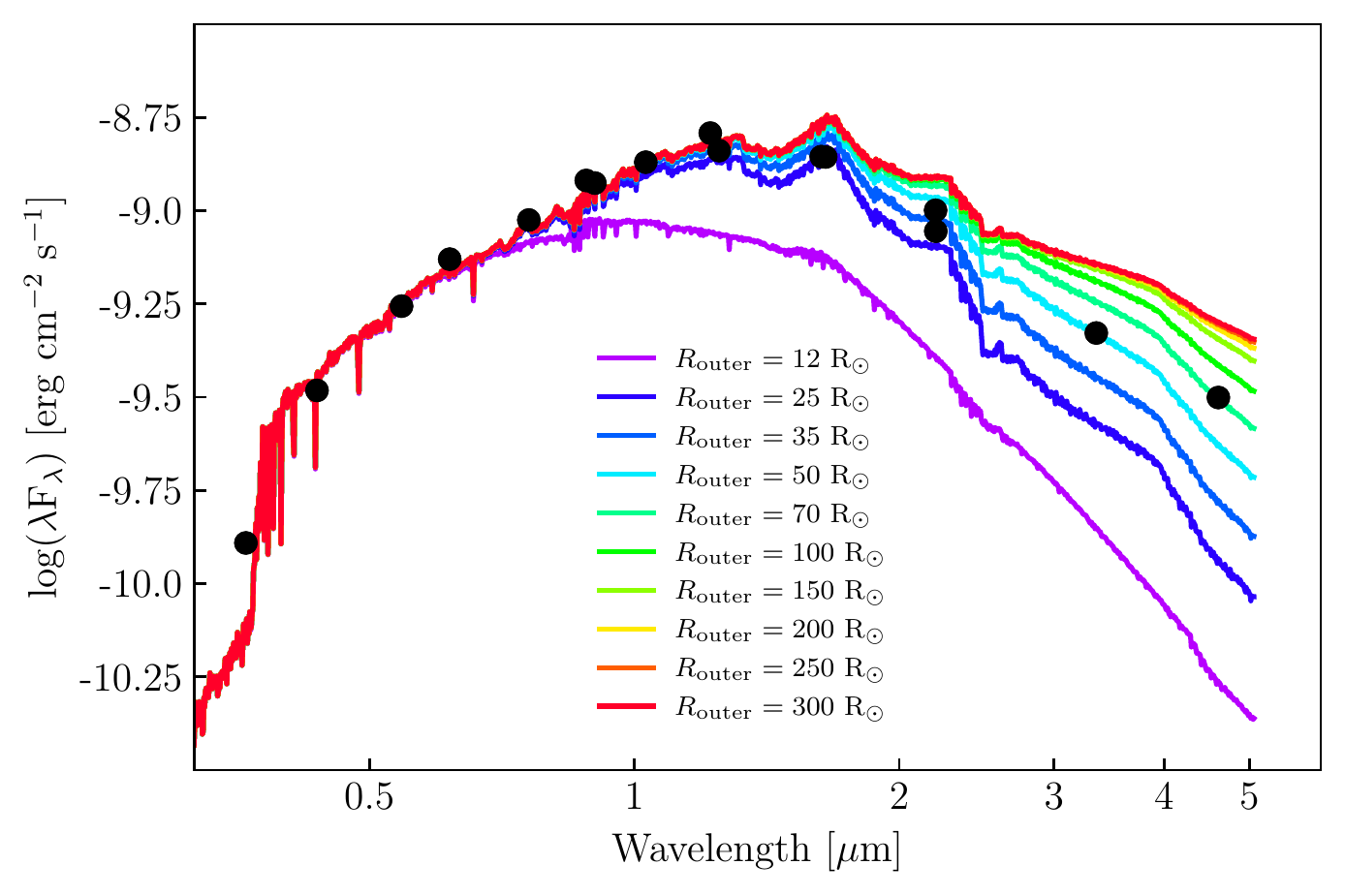}
    \caption{The SED model compared with the photometry from the outburst epoch, with varying values of $R_\mathrm{outer}$. Notice 
    that the longer wavelength data points favor increasingly larger values of $R_\mathrm{outer}$, an inconsistency that can be rectified by adding a passive disk component exterior to the adopted $R_\mathrm{outer}$ for the active accretion disk.
    }
    \label{fig:VaryingRouter}
\end{figure}

To account for the uncertainty in the $i$ reported in \citet{Kospal_ALMA_2021}, we construct a skew-normal prior distribution for $i$. The parameters of the distribution are chosen such that the bounds of the 95 \% confidence interval match the \citet{Kospal_ALMA_2021} likely range. Finally, we fix $R_{\text{outer}}$ to be 45 $\mathrm{R_\odot}$, which is approximately where the model reaches the dust sublimation temperature. We find that when varying $R_{\text{outer}}$, values around 45 $\mathrm{R_\odot}$ are marginally preferred. We demonstrate the effect of varying $R_\mathrm{outer}$ on the SED in Figure \ref{fig:VaryingRouter}. 

\begin{figure}[!htb]
    \centering
    \includegraphics[width=0.98\linewidth]{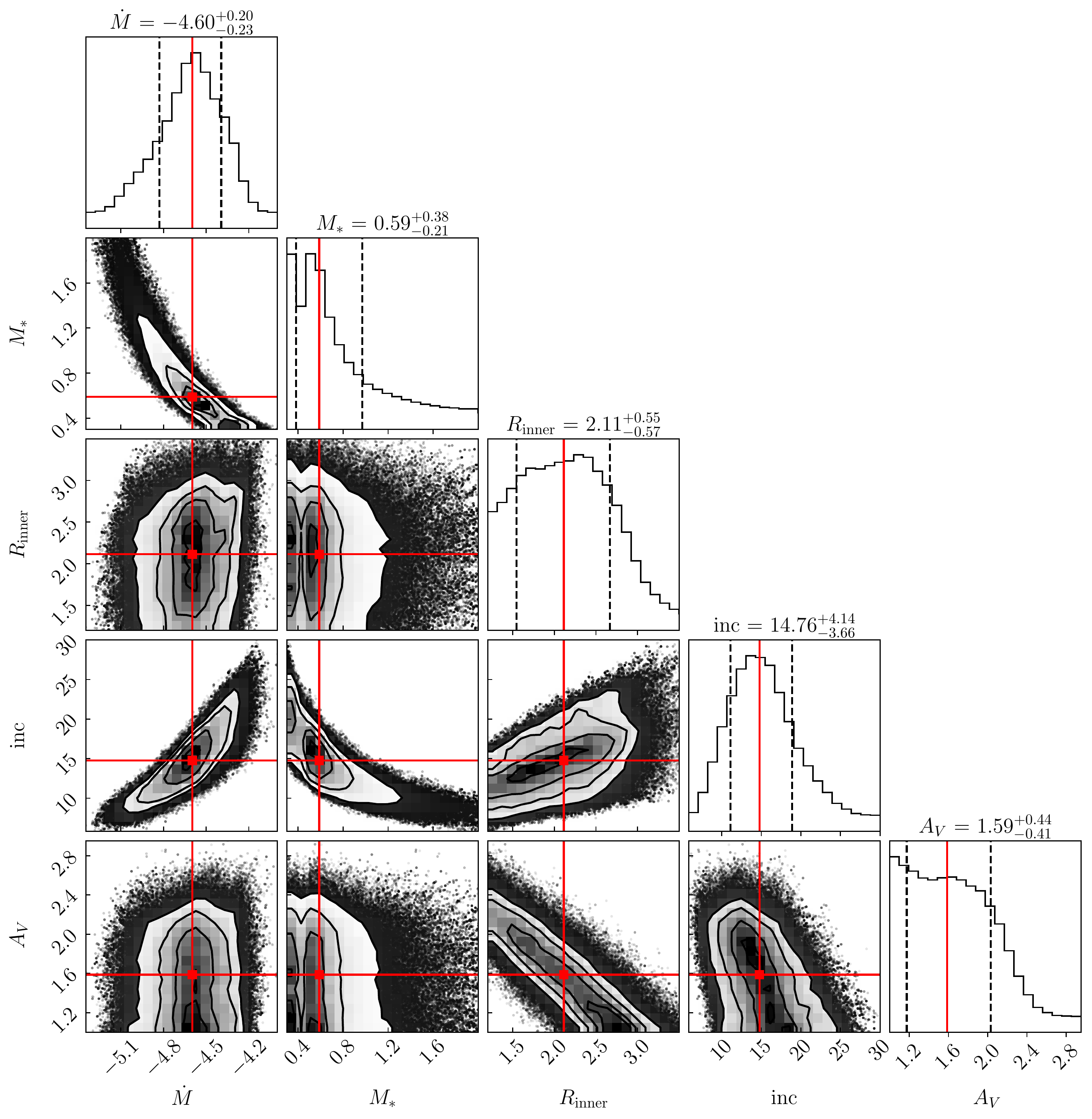}

    \caption{The posteriors from the full SED fit to the outburst SED. All of the parameters are well-constrained, with the exception of $A_V$. }
    \label{fig:full_SEDfit_corner}
\end{figure}

We fit the SED in two major steps via MCMC using the $\mathtt{emcee}$ package \citep{FM_emcee_2013PASP}. In the first step, we allow $M_*$, $\dot{M}$, $R_\mathrm{inner}$, $i$, and $A_V$ to vary , imposing the priors described above. The $\mathtt{corner}$ \citep{corner_FM_2016} plot showing the posterior histograms for this fit is shown in Figure \ref{fig:full_SEDfit_corner}. 

\begin{figure*}[!htb]
    \centering
    \includegraphics[width=0.42\linewidth]{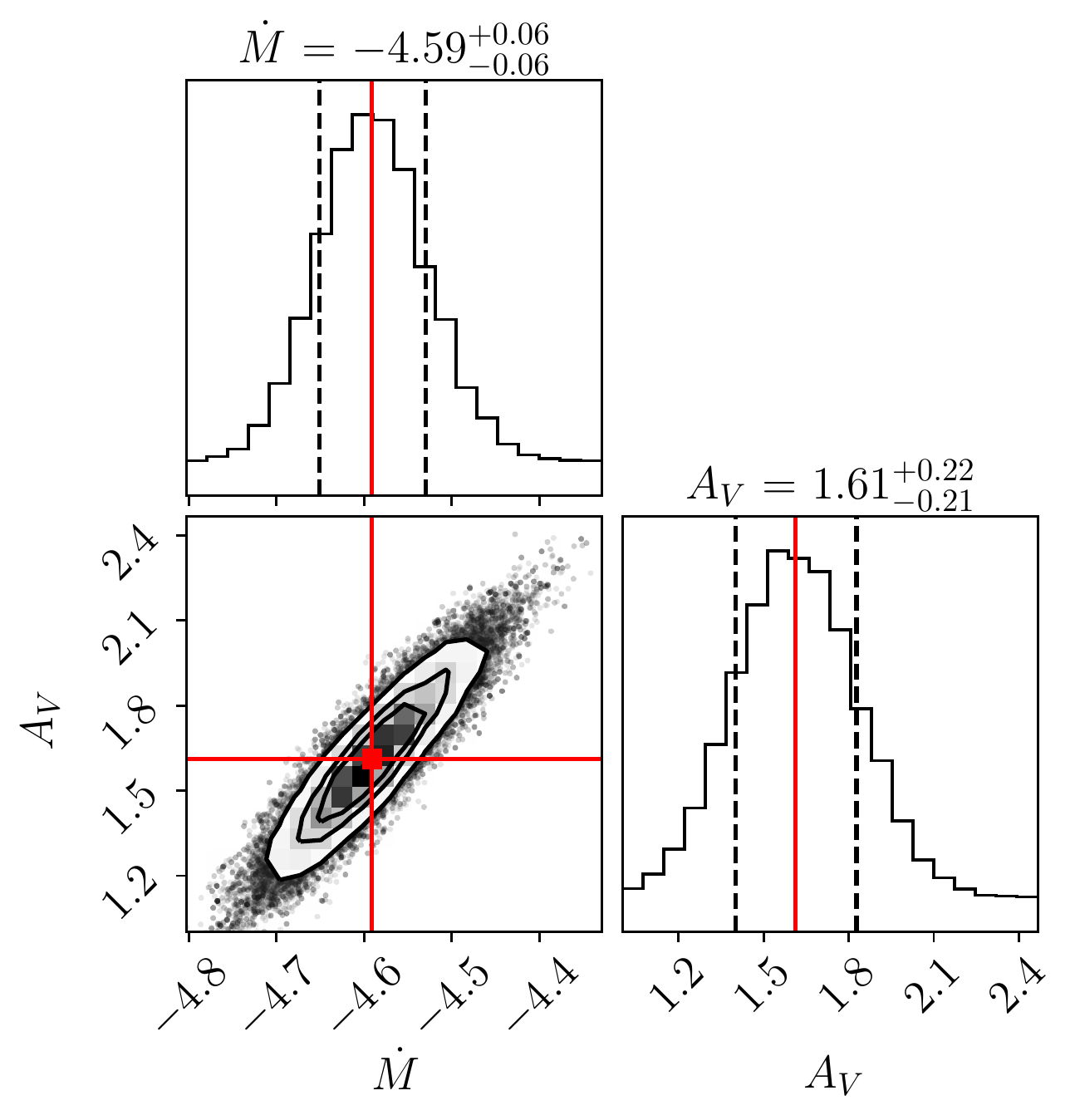}
    \includegraphics[width=0.57\linewidth]{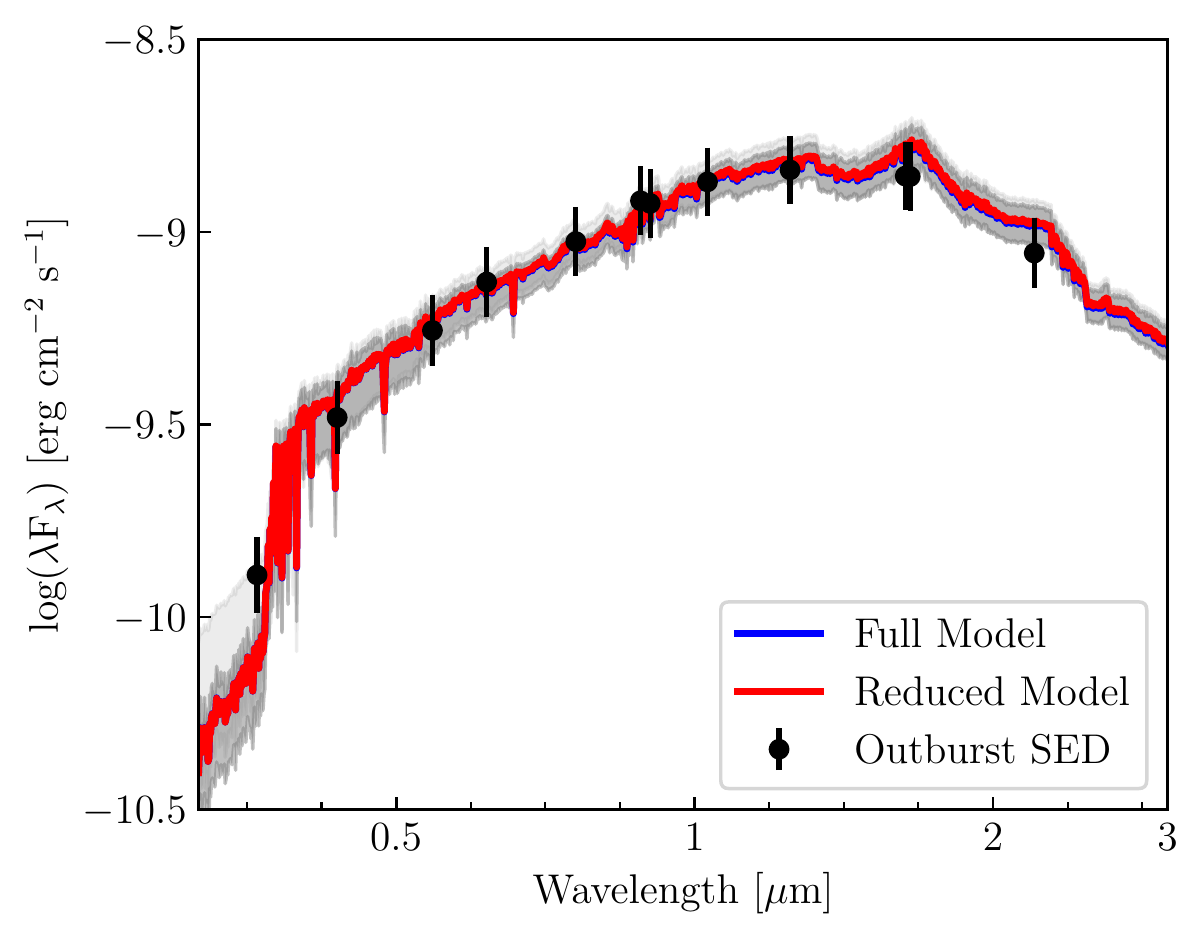}
    \caption{\textbf{Left:} The posteriors for the restricted SED fit, with $M_*=0.59 \ M_\odot$, $i=15^\circ$, and $R_{\text{inner}}=2.11 \ \mathrm{R_\odot}$ fixed to their best-fit values shown in Figure \ref{fig:full_SEDfit_corner}. With the fixed system parameters, the $\dot{M}$ and $A_V$ converge and are consistent with the best-fit values found from the full SED fit. \textbf{Right:} The resulting SED fits, with the full fit shown in blue and the reduced parameter fit shown in red. The good agreement between the two indicates fixing the chosen parameters did not significantly affect the fit. The error bars on the photometry represent 0.01 mag precision.}
    \label{fig:full_SEDFit}
\end{figure*}

The parameters are generally all well-constrained, with the exception of $A_V$. They are also consistent with the estimates from the high dispersion model fits. We therefore fix $M_*=0.59 \ M_\odot$, $R_\mathrm{inner} = 2.11 \ \mathrm{R_\odot}$, and $i = 14.76^\circ$ and in the second step fit only the $\dot{M}$ and $A_V$. The posteriors for this second fit are shown in Figure \ref{fig:full_SEDFit}, along with the SED models from the full and reduced-parameter fit. This second SED model is almost indistinguishable from that produced in the first step of the SED fit and the $\dot{M}$ and $A_V$ best-fit values are consistent with the posteriors in the first step.  

We therefore adopt as the best-fit parameters for the system: $M_* = 0.59 \ M_\odot$, $R_\mathrm{inner} = 2.11 \ \mathrm{R_\odot}$, $\dot{M} = 10^{-4.59} \ M_\odot \ \mathrm{yr}^{-1}$, $i = 15^\circ$, and $A_V = 1.61$. This best-fit model is the one shown in Figures \ref{fig:hires_fits1}, \ref{fig:hires_fits2}, and \ref{fig:hires_fits3}, 
in comparison to the observed high dispersion spectrum  over the orders that were used to compute the initial $\chi^2$ cube. 

Our best-fit system parameters for the disk yield an $L_\mathrm{acc} \sim 100 \ L_\odot$ at outburst. 
For a $0.6 \ M_\odot$, 1 Myr star, the PARSEC isochrones predict that the stellar luminosity should be $L_* = 0.85 \ L_\odot$. 
Thus, the central star contributes less than 1 \% of the total luminosity of the FU Ori object, and thus can be ignored in our modelling procedure.

We take the values of $M_*$, $i$, and $A_V$ as fixed and unchanging, with the justification for adopting a constant $A_V$ provided in \ref{sec:CMDs}. However, we consider the derived $R_\mathrm{inner}$ and $\dot{M}$ parameters as references for the outburst epoch of the system only. In Section \ref{sec:temp} we discuss the temperature evolution of the system, and the effect of varying these two parameters. We assume that during the photometric fade, these are the only two system parameters which change and argue that this is sufficient to explain the variation in both the photometry and spectrophotometry (discussed below) and the high dispersion spectra (Paper II).

\section{Photometric Evidence of Temperature Evolution in the Disk} \label{sec:temp}
The post-outburst fade of V960 Mon can be explained by a decrease in $T_\mathrm{max}$ in the disk, driven by a decrease in $\dot{M}$ and an increase in $R_\mathrm{inner}$. We ignore any stellar contribution to the system, as justified above based on the luminosity arguments.

The early fade of the system is well-sampled by several photometric surveys, enabling us to study the fade in color-magnitude space for several bands. We primarily make use of the AAVSO, ROAD, and ROBOTT photometry for our analysis. We use the photometry to study the color-temperature of the system, estimate the variation in accretion rate during the post-outburst fade, and demonstrate that $R_\mathrm{inner}$ is changing along with the $\dot{M}$ over time. 

\subsection{The color temperature of the system} \label{sec:colorTemp}
\begin{figure}[!htb]
    \centering
    \includegraphics[width=0.98\linewidth]{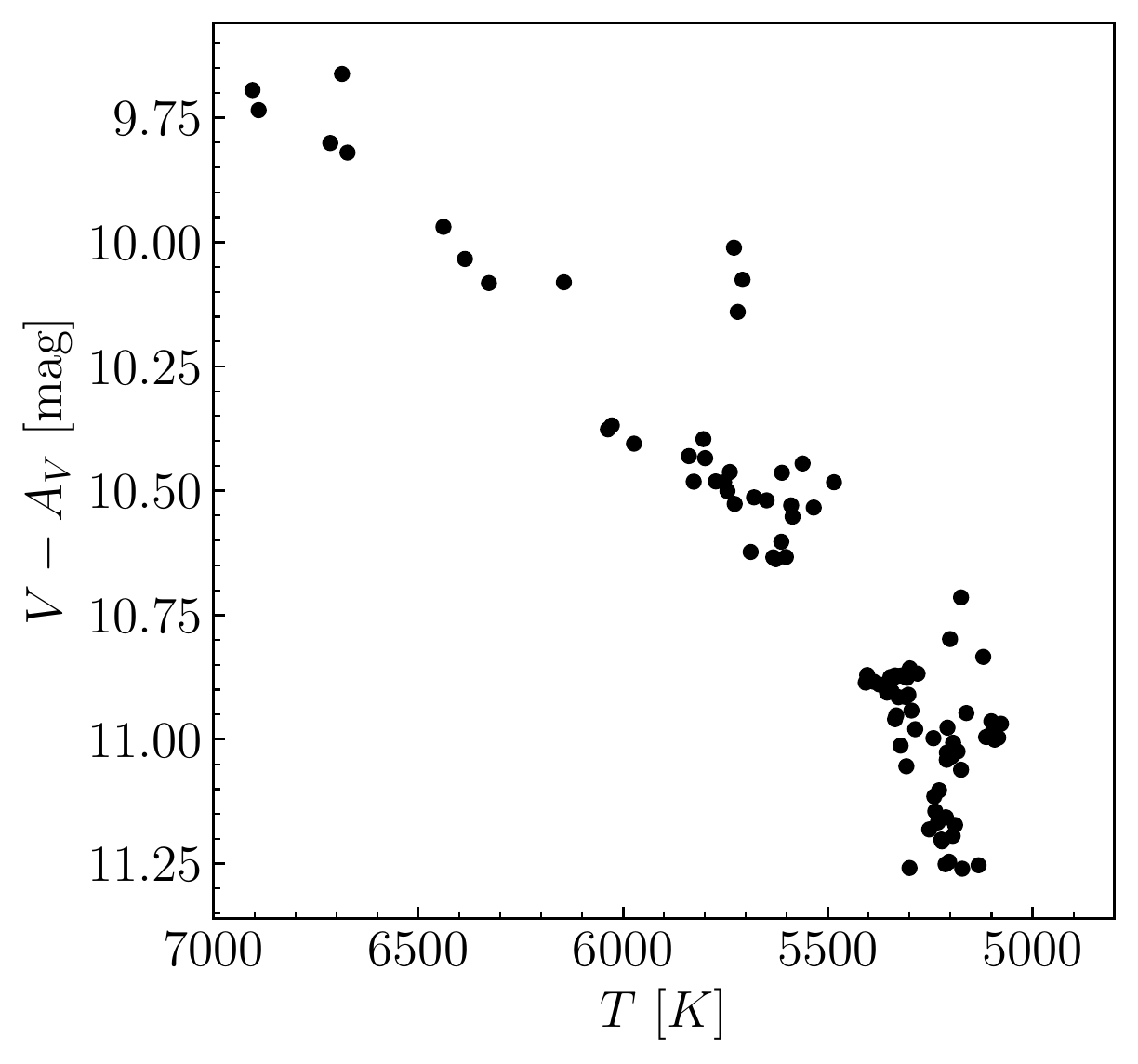}
    \caption{V magnitude of V960 Mon versus color temperature of V960 Mon, computed using the \citet{Ballesteros_2012} conversion from $B-V$ to $T_{\text{eff}}$. $B-V$ is adjusted assuming $E(B-V)=0.53$}
    \label{fig:CMDs}
\end{figure}

We first compute the color-temperature variation in the system, using the AAVSO $B$ and $V$ photometry. We adjust the $V$ magnitudes using the best-fit $A_V = 1.61$ and use $(B-V) - E(B-V)$, assuming the typical interstellar value $R_V = 3.1$ \citep{cardelli_relationship_1989}. We then use the \citet{Ballesteros_2012} calibration to convert the $(B-V) - E(B-V)$ values to temperatures. The color-magnitude diagram with the color temperature conversion is shown in Figure \ref{fig:CMDs}.

The color temperature of the disk at outburst is approximately 6500 K, and cools to 5000 K before reaching the plateau after MJD = 58500. In order to estimate the appropriate disk parameters at later epochs, we explore use of the estimated color temperature evolution and the best-fit system parameters at outburst as a starting point. We can then scale the system parameters (particularly $\dot{M}$ and $R_\mathrm{inner}$) to model the post-outburst disk in a physically motivated and consistent way. 

In T Tauri systems, the innermost accretion boundary is typically estimated to be set by the pressure balance between the magnetic field energy density $P_B = \frac{B^2}{8 \pi}$ and the accretion ram pressure $P_\mathrm{ram} = \frac{1}{2} \rho v^2$. One thus derives the classic equation
\begin{equation} \label{eq:r_inner}
    R_\mathrm{inner} = \left[  \frac{B_*^4 R_*^{12}}{2GM_* \dot{M}^2}  \right]^\frac{1}{7},
\end{equation}
where $R_*$ is the stellar radius and $B_*$ is the stellar surface field strength \citep{Hartmann_review_2016ARA&A}. 

This shows immediately that $R_\mathrm{inner} \propto \dot{M}^{-2/7}$. Recall then that plugging $r = 1.361 \ R_\mathrm{inner}$ into Equation \ref{eq:TProf} gives
\begin{equation}
    T^4_\mathrm{max} = \frac{3 G M_* \dot{M}}{56 \pi \sigma R_\mathrm{inner}^3},
\end{equation}
from which we see $T_\mathrm{max} \propto \dot{M}^\frac{1}{4} R_\mathrm{inner}^{-\frac{3}{4}}$. Using now the fact that $R_\mathrm{inner} \propto \dot{M}^{-\frac{2}{7}}$, we can write $T_\mathrm{max} \propto \dot{M}^\frac{13}{28}$. 

This gives us a way to scale the estimated mass accretion rate at outburst, to that at later epochs by using the relative color temperatures between epochs, while accounting for expected increase in $R_\mathrm{inner}$. We demonstrate good agreement between this model, with an increasing $R_\mathrm{inner}$, and the photometric evolution in Section \ref{sec:CMDs}.

Using $T_{\text{max}} \propto \dot{M}^\frac{13}{28}$ at a given time and the temperature difference between the outburst epoch and each HIRES epoch, we can estimate the $\Delta \dot{M}$ over time as
\begin{equation} \label{eq:temp_profs}
    \frac{T_{\text{epoch}}}{T_{\text{outburst}}} = \left( \frac{\dot{M}_{\text{epoch}}}{\dot{M}_{\text{outburst}}} \right)^{\frac{13}{28}}
\end{equation}

The accretion rates we calculate with Equation \ref{eq:temp_profs} (and resulting $R_\mathrm{inner}$ values)  at each HIRES epoch are reported in Table \ref{tab:SysParams}. The temperature profiles for these system parameters are shown in Figure \ref{fig:tempProfs}. 

\begin{figure}
    \centering
    \includegraphics[width=0.99\linewidth]{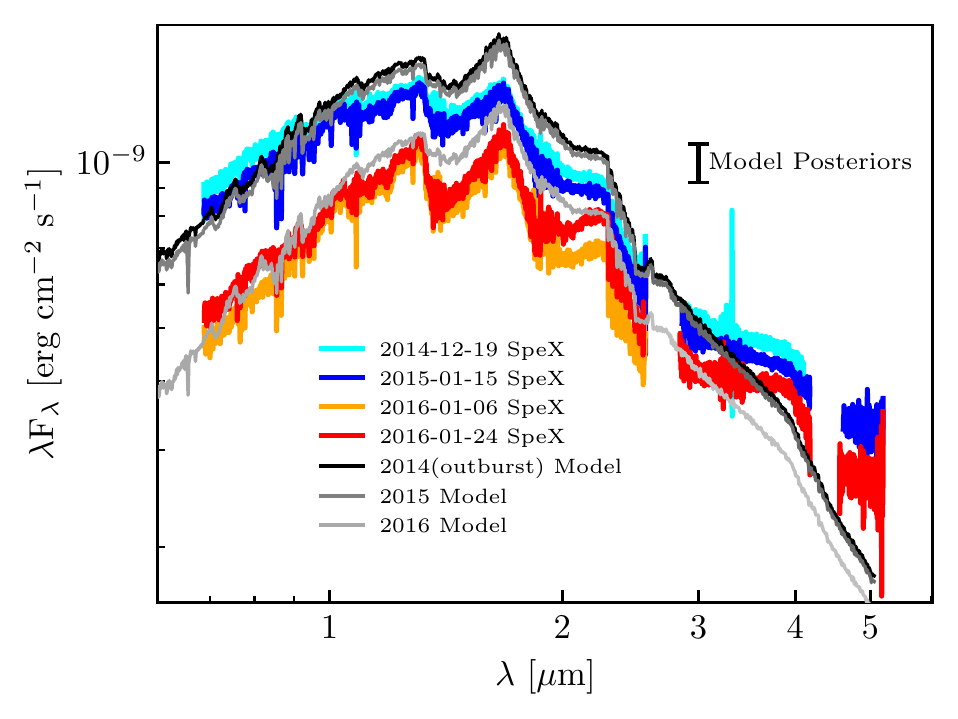}
    \caption{The SpeX spectra compared with the best-fit outburst model and the model generated using the extrapolated system parameters described in Section \ref{sec:colorTemp}. The flux uncertainty is shown with an error bar in the upper right. Overall, the models are in good agreement with the NIR disk emission. However, there are significant discrepancies in between the molecular features in the data and the models, and generally the pure accretion disk models produce poor fits to the L and M bands. The models consistently predict a much stronger $H$ band peak than the data predict, especially at outburst. The discrepancies are further discussed in Section \ref{sec:NIR} and Paper II. }
    \label{fig:SpeX}
\end{figure}

\begin{figure}[!htb]
    \centering
    \includegraphics[width=0.98\linewidth]{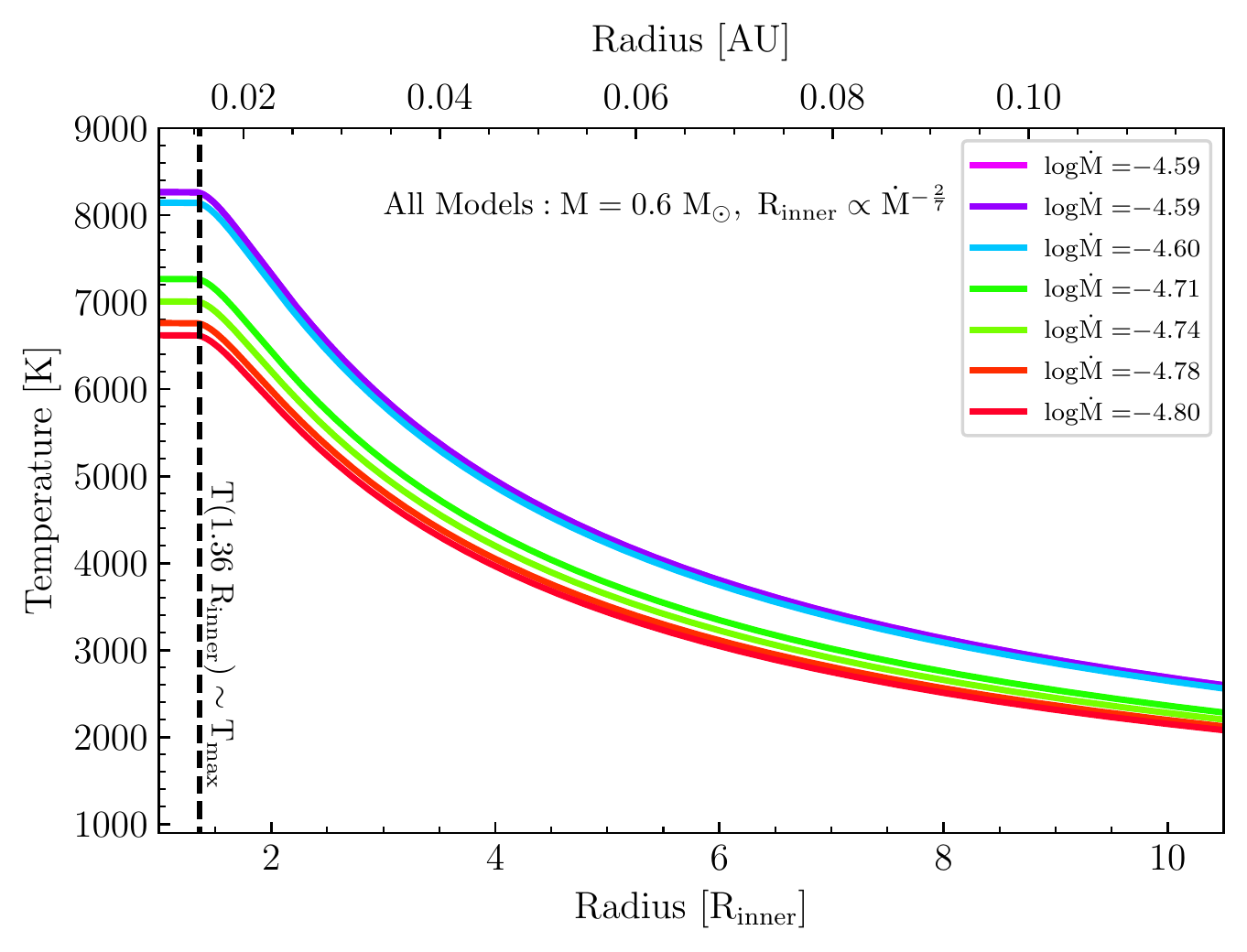}
    \caption{Temperature profiles adopted for the high resolution models of the HIRES epochs. }
    \label{fig:tempProfs}
\end{figure}

We confirm the efficacy of this means of estimating the $\dot{M}$ and $R_\mathrm{inner}$ of the disk for a given epoch by comparing our models with the SpeX spectrophotometry, shown in Figure \ref{fig:SpeX}. The SpeX epoch closest to outburst (19 Dec 2014, cyan spectrum) is consistent with our model using the best-fit outburst parameters derived above. We then scale the $\dot{M}$ and $R_\mathrm{inner}$ to the later epochs and those models are consistent as well. The 24 Jan 2016 epoch (red spectrum) is well-matched by the model that uses $\mathrm{log}\dot{M}=-4.75$ and $R_\mathrm{inner}=2.37 \ R_\odot$. 

We also compare the integrated luminosities of the SEDs and the models in the 0.7-2.5 $\mu$m range and find them to be in good agreement. The earlier two SpeX epochs, have integrated luminosities of 46 (cyan curve) and 49 (blue curve) $L_\odot$ and the two models have luminosities of 49 and 48 $L_\odot$, with discrepancies of $\sim 2 \%$. The later epochs have luminosities of 31 (orange curve) and 36 (red curve) $L_\odot$ and the model has a luminosity of 35 $L_\odot$, in better agreement with the red curve but still within 15 \% of the orange curve.

We address the mismatches between the model and the data in the deep molecular bands of the H and K bands in detail in Paper II.   In Section \ref{sec:NIR} we address the long-wavelength mismatches, in L and M bands of the SpeX spectra, 
which we find can be accounted for by including a passive disk component that reproduces the observed excess 
above the pure-accretion disk model.

Considering the overall evolution of the system, the change in accretion rate is relatively small, with $\dot{M}_\mathrm{2017} = 0.6 \ \dot{M}_\mathrm{outburst}$. If we assumed only that 
\begin{equation}
\frac{\Delta \dot{M}}{\dot{M}_\mathrm{outburst}} \propto \frac{\Delta L_\mathrm{bol}}{L_\mathrm{bol, \ outburst}} \propto 10^{-0.4 \ \Delta V},    
\end{equation}
and used the $\Delta V \sim 1.5$ since outburst, we would find $\dot{M}_\mathrm{2017} \sim 0.25 \ \dot{M}_\mathrm{outburst}$. 


\begin{deluxetable}{cccccc}[!htb]
	\tablecaption{V960 Mon System parameters at each HIRES epoch. Parameters in the first epoch are all best-fit parameters from the procedure described in Section \ref{sec:modelFitting}. For later epochs, $\dot{M}$ and $R_\mathrm{inner}$ are estimated using the color-temperature method in Section \ref{sec:colorTemp}. The parameters shown in italics in later epochs are fixed.  \label{tab:SysParams}}
	\tablewidth{0pt}
	\tablehead{
	    \colhead{Epoch} & \colhead{log$\dot{M}$} & \colhead{$R_\mathrm{inner}$} & \colhead{$M_*$} & \colhead{$A_V$} & \colhead{$i$} \\ 
            \colhead{} & \colhead{($M_\odot$ yr$^{-1}$)} & \colhead{($R_\odot$)} & \colhead{($M_\odot$)} & \colhead{(mag)} & \colhead{($^{\circ}$)} 
	}
    \startdata
    2014-12-09 & -4.59 & 2.11 & 0.59 & 1.61 & 15 \\
    2014-12-10 & -4.59 & 2.11 & \textit{0.59} & \textit{1.61} & \textit{15} \\
    2015-02-09 & -4.60 & 2.13 & \textit{0.59} & \textit{1.61} & \textit{15} \\
    2015-10-27 & -4.72 & 2.29 & \textit{0.59} & \textit{1.61} & \textit{15} \\
    2016-02-02 & -4.75 & 2.35 & \textit{0.59} & \textit{1.61} & \textit{15} \\
    2016-10-14 & -4.78 & 2.40 & \textit{0.59} & \textit{1.61} & \textit{15} \\
    2017-01-13 & -4.81 & 2.44 & \textit{0.59} & \textit{1.61} & \textit{15}     
    \enddata
\end{deluxetable}

\subsection{Color-Magnitude Evolution during the Post-Outburst Fade} \label{sec:CMDs}
In \citet{Hackstein_binary_2015}, the authors point out that the evolution of V960 Mon post-outburst in color-magnitude space does not convincingly follow the $A_V$ vector across all color spaces. In fact, while the initial fade (MJD $<$ 7300) is well-matched by an $A_V$ of 0.5 magnitudes, that amount of extinction significantly overpredicts the amount of fading and reddening in redder color spaces. 

\begin{figure}[!htb]
    \centering
    \includegraphics[width=0.95\linewidth]{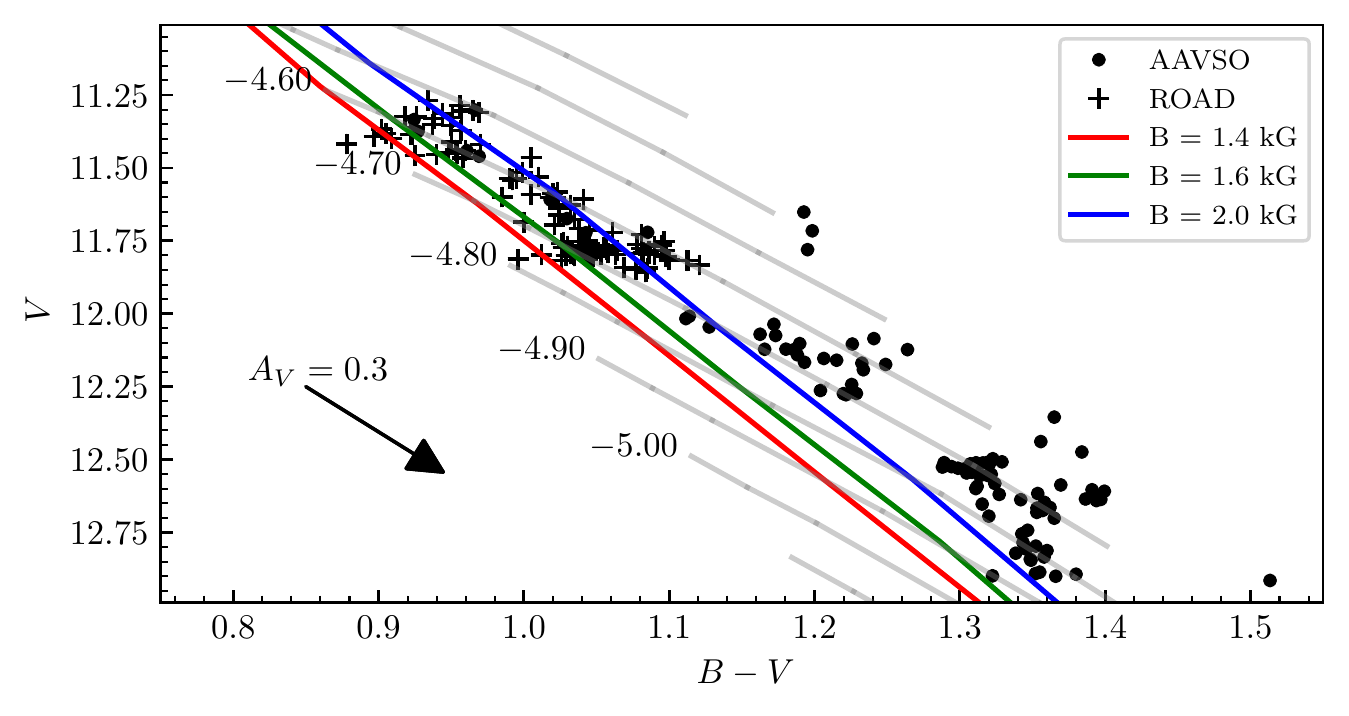}
    \includegraphics[width=0.95\linewidth]{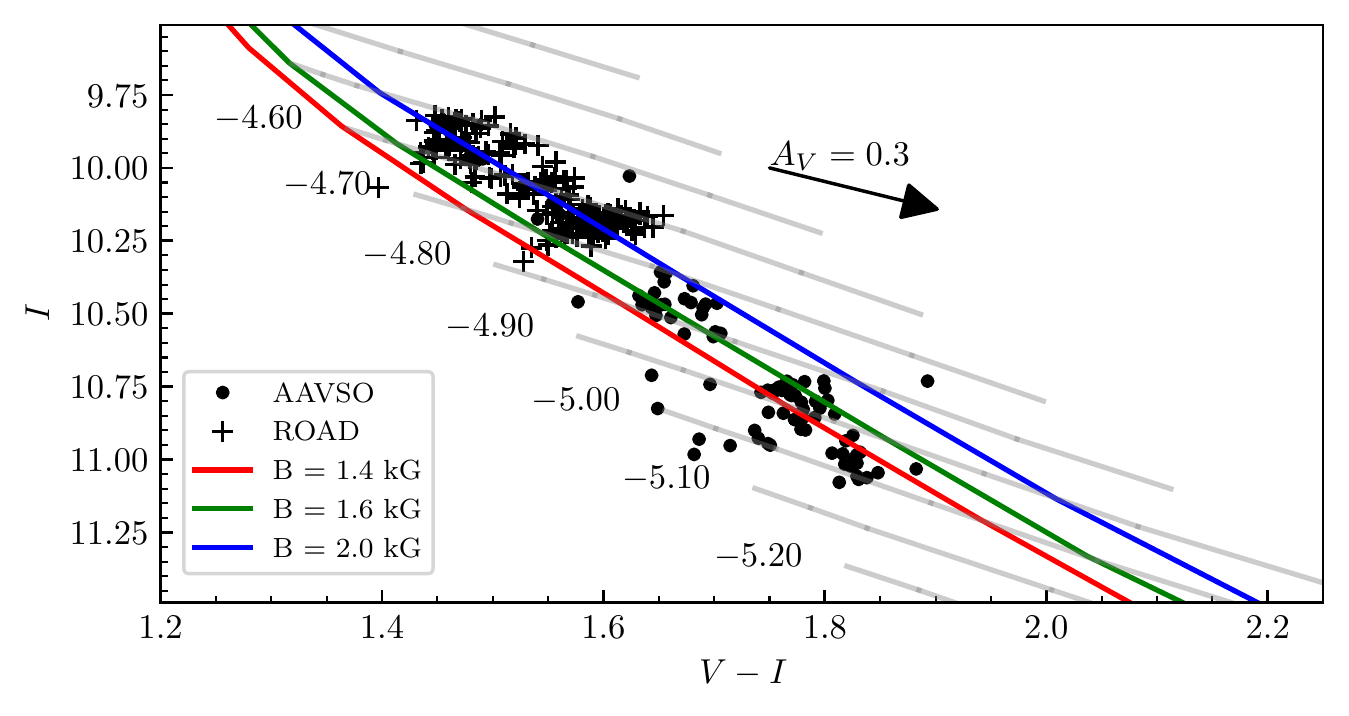}
    \includegraphics[width=0.95\linewidth]{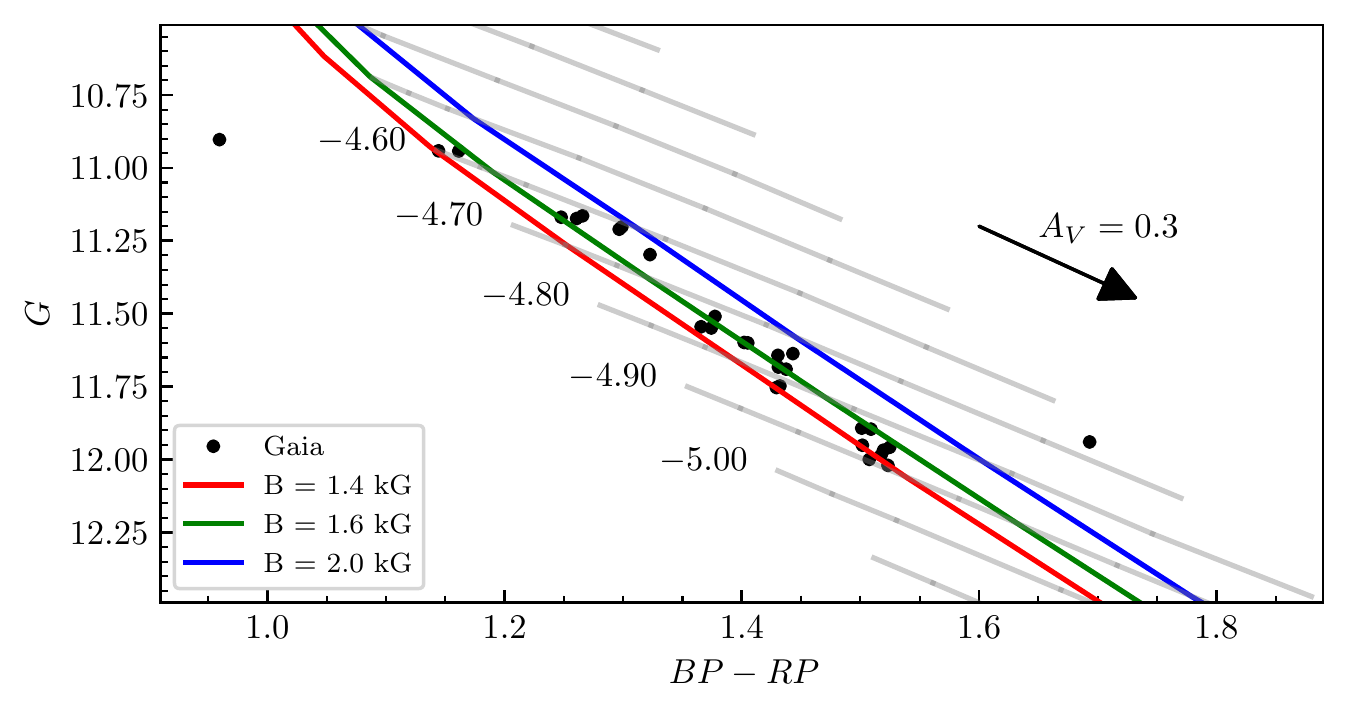}
    \includegraphics[width=0.95\linewidth]{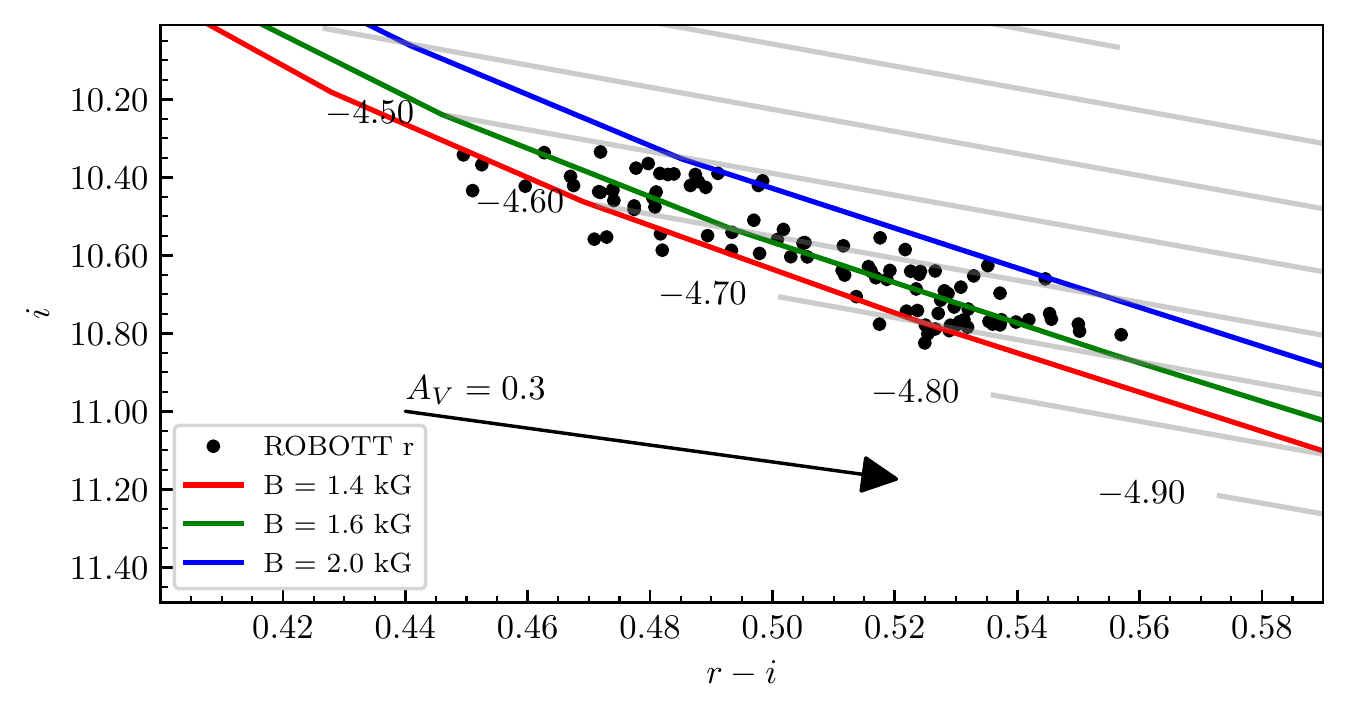}
    \caption{Color magnitude diagrams for V960 Mon using combined photometry from sources shown in Figure \ref{fig:AAVSO_lc}. Model curves are computed with $M_* = 0.6 \ M_\odot$, $A_V = 1.6$ mag, and $d = 1120$ pc \citep{kuhn_comparison_2019}. The grey curves show lines of constant $\dot{M}$ and varying $R_\mathrm{inner}$. The colored curves show lines of $\dot{M}$ varying along with $R_\mathrm{inner}$ according to Equation \ref{eq:r_inner} for different assumed values of $B_*$.  \textbf{Top Two Plots:} $V$ vs $B-V$ and $I$ vs $V-I$ from AAVSO and other sources, along with our models. \textbf{Third Plot:} Gaia $G$ vs $BP - RP$ photometry, converted to $AB$ magnitudes, plotted alongside our models. The Gaia photometry is very well matched by the models.  \textbf{Bottom:} $i$ vs $r-i$ using the ROBOTT (USB) photometry. The ROBOTT photometry spans only the initial steep fade in the lightcurve, prior to JD 2457400. }
    \label{fig:CMDs_models}
\end{figure}

In Figure \ref{fig:CMDs_models}, we show the photometry in four color-magnitude diagrams (CMDs): $(B-V, \ V)$, $(V-I, \ I)$, $(BP-RP, \ G)$, $(r-i, \ i)$. In these four spaces, the necessary amount of extinction to match the evolution of the target differs significantly, as described in \citet{Hackstein_binary_2015}. In fact, not only does the magnitude of extinction fail to match across the four CMDs, the corresponding reddening fails to match the reddening in the photometry. This is especially egregious in the $(V-I, \ I)$ diagram, where the slope of the photometry is much steeper than the slope of the $A_V$ vector. 

We thus conclude that the $A_V$ variation is negligible and we propose that the evolution of the target can be better explained by a change in $T_{\mathrm{max}}$ that is driven by a simultaneous decrease in $\dot{M}$ and an accompanying increase in $R_\mathrm{inner}$. We model several SEDs with different $\dot{M}$ and $R_\mathrm{inner}$ values and compute the magnitudes in several bands using the filter profile package $\mathtt{speclite}$. We use the \texttt{get\_ab\_magnitude} function along with the filter response profiles. We convert AB magnitudes to Vega magnitudes as appropriate. The modeled CMDs are also plotted in Figure \ref{fig:CMDs_models}. The increasing $R_\mathrm{inner}$ reddens the target more rapidly than just the decreasing $\dot{M}$, which is consistent with the photometric evolution in several color spaces. 

The maximum recession rate of $R_\mathrm{inner}$ implied by these models can be estimated by focusing on the rapid fade from MJD 56935 to MJD 57163. The photometry corresponding to this period shows a $\Delta V \sim 0.3$ magnitudes and is well-sampled by both AAVSO and ROAD. The $\dot{M}$ for that epoch would be $10^{-4.75}$ $M_\odot$ yr$^{-1}$, corresponding to $R_\mathrm{inner} \sim 2.4 \ R_\odot$. This is a recession of $0.25 \ \mathrm{R_\odot}$ in 228 days. The implied recession rate, then, is $\sim 8.8$ m s$^{-1}$ or 1870 AU Myr$^{-1}$, This recession rate gives a viscous timescale  $\sim 100$ times greater than those typically associated with T Tauri accretion disks \citep{armitage_accretion_viscosity_2003MNRAS}.

\section{The Passive Disk Contribution to L and M Band} \label{sec:NIR}
As can be seen in Figure \ref{fig:SpeX}, the SpeX spectra (and the W1 and W2 photometry) show a clear L and M band excess that we do not model with our active accretion disk model. Increasing the $R_\mathrm{outer}$ parameter in our model does not account for the excess, and, as discussed in Section \ref{sec:modelFitting}, overestimates the flux in K band. 

We find that adding a passive disk component to the system accounts for the majority of this excess, though matching the SpeX L and M bands precisely is challenging. To compute the contribution of the passive disk, we modify the passive disk prescription given in \citet{Chiang_Goldreich_passive_1997}. We assume a slightly flared disk, with linearly increasing flare angle, such at that a given radius $r$, the height of the disk surface, $h$, is set by $\frac{h}{a} = \eta \ \left( \frac{a}{a_i} \right)^\beta$. 

We estimate the effective temperature, $T_\mathrm{eff}$, of the passive disk reprocessing the disk accretion flux to be given by a version of \citet{Chiang_Goldreich_passive_1997} Equation 1. We generalize the expression to allow for contributions to $T_\mathrm{eff}$ from multiple sources. The modified version is shown below:
\begin{equation}
    T_\mathrm{eff}(a) \approx \left( \frac{\alpha}{2} \right)^{1/4} 
    \left( \sum_\mathrm{i}^\mathrm{N \ sources} \frac{r_i^2}{a^2} \ T_i^4 \right)^{1/4}.
\end{equation}
This allows us to incorporate different annuli of the active disk centered at radii of $r_i$ and effective temperatures $T_i$. $\alpha$ gives the grazing angle at which the incoming radiation strikes the disk. We only model the emission from the two innermost active disk annuli, and since $a >> r_i$, $\alpha$ will not change significantly between the two.   

We compute $\alpha$ assuming the dominant flux is coming from a given annulus of the inner disk, rather than a central star. This new geometry would result in a $\alpha$ given approximately by
\begin{equation}
    \alpha \approx \mathrm{arctan}\left( \frac{h}{a - a_i}  \right) - \mathrm{arctan}\left( \frac{h}{a - r_i}  \right).
\end{equation}
We then use Planck functions $B_\nu(T_\mathrm{eff})$ to model the reprocessed emission from the annuli of the passive disk and integrate the flux. 

\begin{figure}[!htb]
    \centering
    \includegraphics[width=\linewidth]{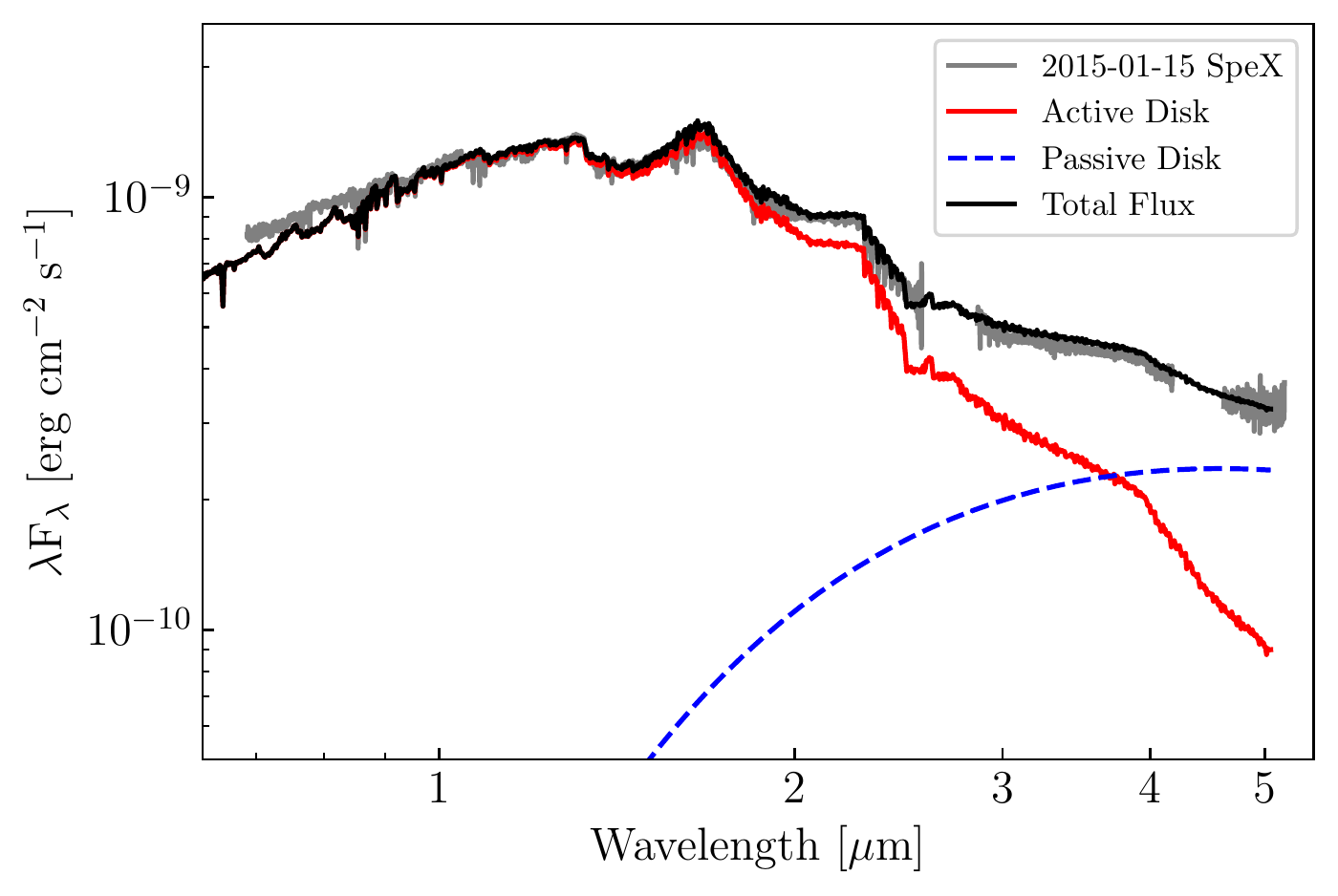}
    \caption{ A model of the system for the 2015-01-15 SpeX epoch, incorporating both an active (red line) and a passive (blue dashed line) disk component. The active disk component is the same model as that shown in Figure \ref{fig:SpeX} for the 2015-01-15 epoch, but computed with $R_\mathrm{outer}=25 \ R_\odot$. The passive disk component uses a flared disk model with a linearly increasing flare angle, described in Section \ref{sec:NIR}. Notice that the addition of the passive disk allows the model to better match the H band molecular features, the K band continuum, and the L and M band excess. }
    \label{fig:PassiveDiskSED}
\end{figure}

We fix the $a_i$ value to be equal to $R_\mathrm{outer}$ of the active component. To best match the K band continuum, we must decrease $R_\mathrm{outer}$ of the active disk from 35 $R_\odot$ to 25 $R_\odot$. Otherwise, the passive disk contribution causes the total model to overestimate the K band flux. We vary the $\eta$ and $\beta$ flaring parameters and find that the flaring law that best matches the decrease in K band, the sudden rise at L band, and the flat M band, is given by $\frac{h}{a} = 0.08 \left( \frac{a}{a_i} \right)^{1.0}$. This is in good agreement with the flaring law found for FU Ori by \citet{lykou_fuori_matisse_2022A&A}, though they find a shallower power law index of 0.13. The final model is shown in Figure \ref{fig:PassiveDiskSED}.

\section{Discussion} \label{sec:discussion}
Our accretion disk model consistently reproduces the evolution of V960 Mon during its post-outburst fade from 2015-2017 in multiple color-magnitude spaces, as well as in the high dispersion spectra
that are discussed in Paper II. 

Combining constraints from the outburst SED and HIRES spectrum, we are able to constrain the system parameters to: $\dot{M} = 10^{-4.59}$  $M_\odot$ yr$^{-1}$, $A_V = 1.61$ mag, $M_*=0.59$  $M_\odot$, $R_\mathrm{inner}=2.11$ $\mathrm{R_\odot}$, and $i=15^\circ$. 

Using the outburst SED we have compiled, the 1120 pc distance to the target, and the best-fit $A_V=1.61$ mag, we compute an $L_\mathrm{bol} = 107 \ L_\odot$. If we compute the $L_\mathrm{acc}$ expected from the system parameters at outburst, we get $L_\mathrm{acc} = 113 \ L_\odot$. This indicates we may be missing some flux, perhaps in the NUV. Integrating our model SED gives $L_\mathrm{bol, model}= 90 \ L_\odot$. However, as shown in Figure~\ref{fig:PassiveDiskSED}, a model of just the active disk misses a significant amount of flux in the NIR. Our inclusion of a passive disk component to the model (see Section \ref{sec:NIR}) accounts for this missing flux by reproducing the L and M band excesses.

\subsection{Comparison with previous results}

Our derived parameters are generally consistent with those found in other literature on the target. Though we use the inclination posteriors reported in \citet{Kospal_ALMA_2021} as a prior in our fits, the SED fit finds a best-fit inclination consistent with theirs. Our best-fit $R_\mathrm{inner}$ value is a bit larger than their reported $R_* = 1.69 \ R_\odot$, but they have a broad 68 \% credible region, extending to $R_* = 6.45 \ R_\odot$ on the larger side. 

The best-fit $A_V$ value is in good agreement with the 1.5 mag measured by \citet{connelley_near-infrared_2018}, and is somewhat smaller than the 2.4 mag estimated by \citet{CarvalhoHillenbrand2022}. This may indicate that the DIB method overestimates $A_V$ values for $E(B-V)>0.5$. 

Our mass is slightly smaller than that estimated for the progenitor object in \citet{kospal_progenitor_2015}. However, using a distance estimate of 1120 pc, it is difficult to reconcile the SED of a 0.6 $M_\odot$ or 0.75 $M_\odot$ object with the progenitor SED compiled by \citet{kospal_progenitor_2015}. 

Though the color temperatures we computed from the \citet{Ballesteros_2012} calibration do not match the $T_\mathrm{max}$ values for the temperature profiles, they do match the temperatures at $r \sim 3 \ R_*$. This is consistent with the result of radiative transfer modeling of FU Ori by \citet{Calvet_FUOriModel_1993}, who find that that the optical continuum arises predominantly from the $r \sim 3 \ R_*$ annulus. Since the color temperatures are derived from $B-V$ colors, we expect the estimated temperature to be that of the $r \sim 3 \ R_*$ region of the disk. 

The outburst amplitudes across the photometric bands are somewhat consistent with the outbursts modeled in \citet{hillenbrand_tony_amps_2022RNAAS} and \citet{Liu_fuorParameterSpace_2022ApJ}. The outburst is slightly greater amplitude at shorter wavelengths ($\Delta B \sim 3$ mag) and smaller in the NIR ($\Delta W2 \sim 2.2$ mag). However, this is not as drastic a difference as that predicted in \citet{hillenbrand_tony_amps_2022RNAAS}. This can be explained by two effects not modeled by \citeauthor{hillenbrand_tony_amps_2022RNAAS}: the pre-outburst accretion shock and the post-outburst variable passive disk. A significant contribution from the magnetospheric accretion shock in the bluer bands (especially $U$) would increase expected pre-outburst $U$ and $B$ magnitude and decrease the outburst amplitude in those bands. 

For the NIR outburst amplitude, we expect the passive disk should be more luminous after outburst, as the incident radiation has both a greater $T_\mathrm{eff}$ and overall luminosity. This will increase the post-outburst passive disk contribution to the SED, increasing the outburst amplitude in the reddest bands, especially W1 and W2. Therefore, it is not entirely surprising that the outburst would have an approximately similar amplitude across all optical/NIR photometric bands.

\subsection{The case for a cooling inner disk}
We find that the exponential post-outburst fade of V960 Mon is well-explained by the inner disk cooling over time. The cooling is primarily driven by a decrease in the disk accretion rate, which in turn causes the $R_\mathrm{inner}$ to increase. 

Our investigation of the evolution of the system in four CMDs shows that a model wherein the disk cools is a better fit to the data than one wherein the extinction increases. A single extinction vector cannot reproduce the slope and magnitude of the color variation of the system in the 4 CMDs consistently. In fact, assuming the $\Delta J \sim 0.9$ mag fade is due to extinction, and using $A_J/A_V=3.55$ \citep{Mathis_extinction_1990ARA&A}, we obtain a corresponding $A_V$ increase of 3.2 mag, which is much greater than the $\Delta V$ observed. 

A cooling inner disk (though accompanied by an increase in $A_V$) is also proposed to explain the long-term fading of V1057 Cyg \citep{Szabo_V1057cyg_2021ApJ}. In their model, \citet{Szabo_V1057cyg_2021ApJ} vary the product $M_* \dot{M}$ and $A_V$ over time, finding that an exponential decrease in $\dot{M}$ and a factor 2 increase in $A_V$ is consistent with the SED evolution of the target over 50 years. It is possible that the increase in $A_V$ might be substituted for an increase in $R_\mathrm{inner}$ in this system to better fit the data. The slopes of the CMDs shown in their Figure 7 are inconsistent with either $\dot{M}$ or $A_V$ vectors. However, comparing their d$V$/d$(V-I)$, the overall fading and reddening in the CMD is approximately consistent with the d$V$/d$(V-I)$ we obtain from our model. Rather than varying $A_V$, varying $R_\mathrm{inner}$ over time could better explain the slope seen in the CMDs. 

Ultimately, a cooling inner disk is more consistent with the photometric evolution of systems like V960 Mon and V1057 Cyg than only increasing $A_V$. We discuss the issue of varying $R_\mathrm{inner}$ in Section \ref{sec:varyingRinner}, supported by evidence presented in Section \ref{sec:CMDs}.

\subsection{The case for an increasing $R_\mathrm{inner}$} \label{sec:varyingRinner}
Using the time-evolution of the object in color-magnitude space and the fixed $\dot{M}_{\text{outburst}} = 10^{-4.59} \ M_\odot$ yr$^{-1}$, we have estimated the $\dot{M}$ as a function of time for the system. Assuming only $\dot{M}$ varies over time, we find $\dot{M} = 10^{-5.05} \ M_\odot$ yr$^{-1}$ for the 2017 HIRES epoch. This represents a maximum decrease of a factor of 2.9 in the accretion rate since outburst. 

For the February 2016 epoch, we find that $R_\mathrm{inner}=2.4 \ R_\odot$ and $\dot{M} = 10^{-4.75} \ M_\odot$ yr$^{-1}$ fits the SpeX spectrum well (See Section \ref{sec:colorTemp}). For this same epoch, varying only $\dot{M}$ gives an estimate of $10^{-4.89} \ M_\odot$ yr$^{-1}$, which is a poor fit to the SpeX spectrum. This difference indicates that varying only $\dot{M}$ causes us to underestimate it at later times. 

The question that arises then is whether our use of Equation \ref{eq:r_inner} is physically reasonable in this system. For a 0.6 $M_\odot$, 2.1 $R_\odot$ star \citep[corresponding to a $\sim 0.7$ Myr central star; ][]{Baraffe_isochrones_2015A&A}, we can solve for the necessary $B_*$ such that $R_\mathrm{inner} = R_*$. Doing this gives a value of $\sim 1.4$ kG, well within the range of, and in fact at bit smaller than, typical T Tauri star surface field strengths \citep{Johns-Krull_TTauriBFields_2007ApJ}.

Though this $B_*$ value is slightly smaller than that of the track along which the photometry evolves in Figure \ref{fig:CMDs}, there are factors of order unity, such as the partition of pressure support between gas pressure and magnetic field pressure, which might contribute to this difference. Another possible contribution to the pressure support in the system is that of the magnetic field of the disk, which may be comparable to the $B_*$ value we estimate here \citep{Donati_FUOri_2005Natur, Zhu_outburst_FUOri_2020MNRAS}. 

We find that this relation between $\dot{M}$ and $R_\mathrm{inner}$ is consistent with what we observe in the high dispersion spectra, as presented in Paper II.  Of note is the fact that there are many subtleties in the spectral data, with line strength and line broadening that is sometimes nonintuitive in terms of behavior with excitation potential or wavelength, but in fact can be well-explained by the disk model.

\section{Summary and Conclusion} \label{sec:conclusion}
We have demonstrated that a single accretion disk model can be used to fit the SED and high dispersion spectrum of V960 Mon at outburst. The model can then be modified by varying $\dot{M}$ and $R_\mathrm{inner}$ in physically consistent ways to reproduce the photometric evolution of the system during its post-outburst fade. 

In summary:
\begin{enumerate}
    \item We have modified the accretion disk model described in \citet{Rodriguez_model_2022} to more accurately and rapidly produce both SEDs and high dispersion model spectra. 
    \item We introduced a hybrid technique for using high dispersion spectra of FU Ori objects to inform priors on contemporaneous SED fits of the targets. The technique produces best-fit system parameters near the lightcurve peak of $\dot{M} = 10^{-4.59}$  $M_\odot$ yr$^{-1}$, $A_V = 1.61$ mag, $M_*=0.59$  $M_\odot$, $R_\mathrm{inner}=2.11$ $\mathrm{R_\odot}$, and $i=15^\circ$. We find $R_\mathrm{outer} = 45 \ R_\odot$ produces models most consistent with the $K$ band photometry.
    The $L_\mathrm{acc}$ computed from these parameters is 113 $L_\odot$, in good agreement with the $L_\mathrm{bol} = 107 \ L_\odot$ computed by integrating the outburst epoch SED. 
    \item We used the color-temperature evolution of the system, along with the derived relation $T_\mathrm{eff} \propto \dot{M}^{-13/28}$, to trace the photometric evolution of the system. To better match the evolution as seen in the CMDs, we allowed $R_\mathrm{inner}$ to vary, assuming a justified $R_\mathrm{inner} \propto \dot{M}^{-2/7}$. We assume all other system parameters are fixed during this time.
    \item We investigated the contribution of a passive disk to emission at $L$ and $M$ band in the NIR, finding a simple blackbody representation produces a consistent model, even improving the agreement with the H and K band. The K, L, and M band are well-matched by reducing the $R_\mathrm{outer}$ of the active component to 25 $R_\odot$ and introducing a passive component with a flaring law of $\frac{h}{a} = 0.08 \left( \frac{a}{a_i} \right)$. 
\end{enumerate}

We study the evolution of the system at high dispersion in Paper II. In future work, we will expand this technique to other well-known FU Ori objects. For other objects that show an exponential fade, a cooling inner disk driven by a decrease in $\dot{M}$ and increase in $R_\mathrm{inner}$ may be the best explanation for the systems' fading.

\section{Acknowledgements}
We thank Antonio Rodriguez for insightful conversations and comments. 

We thank the American Association of Variable Star Observers for their dedicated sampling of the post-outburst lightcurve of this target in multiple bands. 


We thank the referee for a careful review of our work.

\facilities{AAVSO, Keck:I (HIRES), IRTF (SpeX), Gaia}

\bibliography{references}{}
\bibliographystyle{aasjournal}

\appendix 

\restartappendixnumbering


\section{Photometry used in Figure 1}
A significant fraction of the photometry used in Figure \ref{fig:AAVSO_lc} can be found on a variety of public databases and in \citet{Hackstein_binary_2015}. We present in Table \ref{tab:phot} the photometry used in Figure \ref{fig:AAVSO_lc}, and throughout this work, that is not already published. All of the photometry from RoBoTT, ROAD, and IRIS can be found in the CDS table associated with \citet{Hackstein_binary_2015}.

\begin{deluxetable}{ccccc}[!htb]
	\tablecaption{Photometry from AAVSO, Gaia, Gattini, and WISE, shown in Figure \ref{fig:AAVSO_lc}. The full, machine-readable version of the table is available in the online journal  \label{tab:phot}}
	\tablewidth{0pt}
	\tablehead{
	    \colhead{Epoch} & \colhead{Flux} & \colhead{Flux Uncertainty} & \colhead{Source/Facility} & \colhead{Band} \\ 
            \colhead{(JD - 2450000.5)} & \colhead{(mag)} & \colhead{(mag)} & \colhead{} & \colhead{} 
	}
    \startdata
    7010.51 & 12.25 & 0.01 &  AAVSO  &  B  \\ 
    7010.51 & 12.25 & 0.01 &  AAVSO  &  B  \\ 
    7012.10 & 12.27 & 0.01 &  AAVSO  &  B  \\ 
    7012.11 & 12.28 & 0.01 &  AAVSO  &  B  \\ 
    7013.10 & 12.27 & 0.01 &  AAVSO  &  B  \\ 
    7013.10 & 12.29 & 0.01 &  AAVSO  &  B  \\ 
    7014.10 & 12.28 & 0.01 &  AAVSO  &  B  \\ 
    7014.10 & 12.27 & 0.01 &  AAVSO  &  B  \\ 
    7015.10 & 12.24 & 0.01 &  AAVSO  &  B  \\ 
    7015.10 & 12.27 & 0.01 &  AAVSO  &  B  \\
    7016.09 & 12.28 & 0.01 &  AAVSO  &  B  \\ 
    7016.09 & 12.27 & 0.01 &  AAVSO  &  B  \\
    $\cdots$ & $\cdots$ & $\cdots$ & $\cdots$ & $\cdots$ 
    \enddata
\end{deluxetable}

\end{document}